\documentclass{ws-rv975x65}  

\newcommand{\Dash}{\boldmath $-$}  

\newcommand{\al}{\alpha}
\newcommand{\be}{\beta}

\newcommand{\De}{\Delta}

\newcommand{\eps}{\epsilon}

\newcommand{\ka}{\kappa}

\newcommand{\beq}{\begin{equation}}
\newcommand{\eeq}{\end{equation}}
\newcommand{\ba}{\begin{array}}
\newcommand{\ea}{\end{array}}
\newcommand{\bea}{\begin{eqnarray}}
\newcommand{\eea}{\end{eqnarray}}
\newcommand{\ben}{\begin{enumerate}} 
\newcommand{\een}{\end{enumerate}}
\newcommand{\bc}{\begin{center}}
\newcommand{\ec}{\end{center}}

\newcommand{\ha}{\frac{1}{2}}
\newcommand{\rr}{{\bf r}}
\newcommand{\q}[2]{ {\bf q}_{#1}^{#2}}
\newcommand{\setq}[2]{\{{\bf q}_{#1}^{#2}\}}
\newcommand{\coleps}{\epsilon_{I\alpha\beta}}
\newcommand{\flaeps}{\epsilon_{Iij}}

\newcommand{\cf}{\alpha i,\beta j}

\newcommand{\dmu}{\delta\mu}

\renewcommand{\>}{\rangle} 
\newcommand{\txt}{\textstyle}

\newcommand\eqn[1]{(\ref{#1})}      

\newcommand{\half} {{\txt \frac{1}{2}}}
\newcommand{\third}{{\txt \frac{1}{3}}}

\newcommand{\twothirds}{{\txt \frac{2}{3}}}
\newcommand{\MeV}{{\rm MeV}} 

\newlength{\vpad}
\usepackage[dvips]{color}

\usepackage[normalem]{ulem}  

\begin{document}

\setcounter{chapter}{0}

\chapter[COLOR SUPERCONDUCTIVITY IN DENSE
QUARK MATTER]{COLOR SUPERCONDUCTIVITY IN 
DENSE, BUT NOT ASYMPTOTICALLY DENSE, 
DENSE QUARK MATTER}

\author{Mark Alford}
\address{ Physics Department, Washington University \\
Saint Louis, MO 63130, USA}
\author{Krishna Rajagopal}
\address{Center for Theoretical Physics, 
Massachusetts Institute of Technology\\ 
 Cambridge, MA 02139 USA\\ 
Nuclear Science Division, Lawrence Berkeley National Laboratory\\ 
Berkeley, CA 94720, USA }

\begin{abstract}
At ultra-high density, matter is expected to form a degenerate Fermi
gas of quarks in which there is a condensate of Cooper pairs of quarks
near the Fermi surface: color superconductivity. In this chapter
we review some of the underlying physics, and discuss outstanding
questions about the phase structure of ultra-dense quark matter.  We
then focus on describing recent results on the crystalline color
superconducting phase that may be the preferred form of cold, dense
but not asymptotically dense, three-flavor quark matter.  The gap
parameter and free energy for this phase have recently been evaluated
within a Ginzburg-Landau approximation for many candidate crystal
structures.  We describe the two that are most favorable. The
robustness of these phases results in their being favored over wide
ranges of density.  However, it also implies that the Ginzburg-Landau
approximation is not quantitatively reliable.  We describe qualitative
insights into what makes a crystal structure favorable which can be
used to winnow the possibilities. We close with a look ahead at the
calculations that remain to be done in order to make quantitative
contact with observations of compact stars.
\end{abstract}

\section{Introduction}
\label{sec:intro}

The exploration of the phase diagram of matter at ultra-high temperature
or density is an area of great interest and activity, both on the
experimental and theoretical fronts. Heavy-ion colliders such as
the SPS at CERN and RHIC at Brookhaven have probed the high-temperature
region, creating and studying the properties of quark matter with very high 
energy density and very low baryon number
density similar to the fluid which filled the universe for 
the first microseconds after the big bang.
In this paper we discuss a different part of
the phase diagram, the low-temperature high-density region. 
Here there are
as yet no experimental constraints, and our goal is
to understand the properties of matter predicted by QCD well enough
to be able to use astronomical observations of neutron stars to
learn whether these densest objects in the current universe contain
quark matter in their core. 
We expect cold, dense, matter to exist in
phases characterized by Cooper pairing of quarks, i.e.
 color superconductivity, driven by
the Bardeen-Cooper-Schrieffer (BCS)\cite{BCS} 
mechanism. The BCS mechanism operates when there is
an attractive interaction between fermions at a Fermi surface.
The QCD quark-quark interaction is strong, and is attractive
in many channels, so we expect cold dense quark matter to {\em generically}
exhibit color superconductivity.
Moreover, quarks, unlike electrons, have color and flavor as well as spin
degrees of freedom, so many different patterns of pairing are possible.
This leads us to expect a rich phase structure
in matter beyond nuclear density.

Calculations using a variety of methods agree that at sufficiently
high density, the favored phase is color-flavor-locked (CFL)
color-superconducting quark matter\cite{Alford:1998mk} (for reviews,
see Ref.~\refcite{Reviews}).  However, there is still uncertainty over
the nature of the next phase down in density. Previous
work\cite{Alford:2003fq,Alford:2004hz} had suggested that when the
density drops low enough so that the mass of the strange quark can no
longer be neglected, there is a continuous phase transition from the
CFL phase to a new gapless CFL (gCFL) phase, which could lead to
observable consequences if it occurred in the cores of neutron
stars.\cite{Alford:2004zr} However, it now appears that some of the
gluons in the gCFL phase have imaginary Meissner masses, indicating an
instability towards a lower-energy
phase.\cite{Huang:2004bg,Casalbuoni:2004tb,Giannakis:2004pf,Alford:2005qw,Huang:2005pv,Fukushima:2005cm,Gorbar:2005rx,Kryjevski:2005qq,Gorbar:2006up,Iida:2006df,Fukushima:2006su,Gerhold:2006dt}
Analysis in the vicinity of the unstable gCFL phase cannot determine
the nature of the lower-energy phase that resolves the
instability. However, the instability is telling us that the system
can lower its energy by turning on currents, suggesting that the
crystalline color superconducting phase,\cite{Alford:2000ze} in which
the condensate is modulated in space in a way that can be thought of
as a sum of counterpropagating currents, is a strong candidate.


\section{Review of color superconductivity}

\subsection{Color superconductivity}

The fact that QCD is asymptotically free implies that
at sufficiently high density and low temperature,
there is a  Fermi surface of weakly-interacting quarks. 
The interaction between these quarks is
certainly attractive in some channels
(quarks bind together to form baryons), so we expect the formation
of a condensate of Cooper pairs.
We can see this by considering the grand canonical potential
$F= E-\mu N$, where $E$ is
the total energy of the system, $\mu$ is the chemical potential, and
$N$ is the number of quarks. The Fermi surface is defined by a
Fermi energy $E_F=\mu$, at which the free energy is minimized, so
adding or subtracting a single particle costs zero free energy. 
Now switch on a weak attractive interaction.
It costs no free energy to
add a pair of particles (or holes), and if they have the
right quantum numbers then the attractive
interaction between them will lower the free energy of the system.
Many such pairs will therefore
be created in the modes near the Fermi surface, and these pairs,
being bosonic, will form a condensate. The ground state will be a
superposition of states with all numbers of pairs, breaking the
fermion number symmetry. 

A pair of quarks cannot be a color singlet, so
the resulting condensate will break the local color symmetry
$SU(3)_{\rm color}$.  The formation of a condensate
of Cooper pairs of quarks is therefore called ``color superconductivity''.
The condensate plays the same role here as the Higgs condensate
does in the standard model: the color-superconducting phase
can be thought of as the Higgs phase of QCD.

\subsection{Highest density: Color-flavor locking (CFL)}
\label{sec:CFL}

It is by now well-established that at sufficiently high densities,
where the up, down and strange quarks can be treated
on an equal footing and the disruptive effects of the
strange quark mass can be neglected, quark matter
is in the color-flavor locked (CFL) phase, in which
quarks of all three colors and all three flavors form
conventional Cooper pairs with zero total momentum,
and all fermionic
excitations are gapped, with the gap parameter 
$\Delta_0\sim 10-100$~MeV.\cite{Alford:1998mk,Reviews}
This has been confirmed by both NJL\cite{Alford:1998mk,Alford:1999pa,Reviews} and 
gluon-mediated interaction calculations.\cite{Schafer:1999fe,Shovkovy:1999mr,Evans:1999at,Reviews}
The CFL pairing pattern is\cite{Alford:1998mk} 
\begin{equation}
\begin{array}{c}
\langle q^\alpha_i C \gamma_5 q^\beta_j \rangle^{\phantom\dagger}
=\Delta_0(\kappa+1)\delta^\alpha_i\delta^\beta_j 
+ \Delta_0(\kappa-1) \delta^\alpha_j\delta^\beta_i 
 = \Delta_0\eps^{\al\be N}\eps_{ij N} + \Delta_0\ka(\cdots)  \\[2ex]
 {[SU(3)_{\rm color}]}
 \times \underbrace{SU(3)_L \times SU(3)_R}_{\displaystyle\supset [U(1)_Q]}
 \times U(1)_B 
 \to \underbrace{SU(3)_{C+L+R}}_{\displaystyle\supset [U(1)_{{\tilde Q} }]} 
  \times \mathbb{Z}_2
\end{array}
\label{CFLcond}
\end{equation}
Color indices $\alpha,\beta$ and flavor indices $i,j$ run from 1 to 3,
Dirac indices are suppressed,
and $C$ is the Dirac charge-conjugation matrix.
The term multiplied by $\kappa$ corresponds to pairing in the
$({\bf 6}_S,{\bf 6}_S)$, which
although not energetically favored
breaks no additional symmetries and so
$\kappa$ is in general small but nonzero.\cite{Alford:1998mk,Schafer:1999fe,Shovkovy:1999mr,Pisarski:1999cn}
The Kronecker $\delta$'s connect
color indices with flavor indices, so that the condensate is not
invariant under color rotations, nor under flavor rotations,
but only under simultaneous, equal and opposite, color and flavor
rotations. Since color is only a vector symmetry, this
condensate is only invariant under vector flavor+color rotations, and
breaks chiral symmetry. The features of the CFL pattern of condensation 
are\cite{Alford:1998mk} 
\begin{itemize}
\setlength{\itemsep}{-0.7\parsep}
\item[\Dash] The color gauge group is completely broken. All eight gluons
become massive. This ensures that there are no infrared divergences
associated with gluon propagators.
\item[\Dash]
All the quark modes are gapped. The nine quasiquarks 
(three colors times three flavors) fall into an ${\bf 8} \oplus {\bf 1}$
of the unbroken global $SU(3)$. Neglecting corrections of order $\kappa$
(see \eqn{CFLcond})
the octet and singlet quasiquarks have gap parameter $\Delta_0$ and 
$2\Delta_0$ respectively.
\item[\Dash] 
A rotated electromagnetic gauge symmetry (generated
by ``${\tilde Q} $'')
survives unbroken. The single remaining massless gauge
boson is a linear combination
of the original photon and one of the gluons.
\item[\Dash] Two global symmetries are broken,
the chiral symmetry and baryon number, so there are two 
gauge-invariant order parameters
that distinguish the CFL phase from the QGP,
and corresponding Goldstone bosons which are long-wavelength
disturbances of the order parameter. 
When the light quark mass is nonzero it explicitly breaks
the chiral symmetry and gives a mass
to the chiral Goldstone octet, but the CFL phase is still
a superfluid, distinguished by its spontaneous baryon number breaking.
%
\end{itemize}

\subsection{Less dense quark matter: stresses on the CFL phase}

The CFL phase is characterized by pairing between different flavors
and different colors of quarks. We can easily understand why this
is to be expected. Firstly, the QCD interaction between two
quarks is most attractive in the channel that
is antisymmetric in color (the $\bar{\bf 3}$). Secondly, pairing
tends to be stronger in channels that do not break rotational 
symmetry,\cite{IwaIwa,Schafer:2000tw,Buballa:2002wy,Alford:2002rz,Schmitt:2002sc,Schmitt:2004et},
so we expect the pairs to be spin singlets,
i.e.~antisymmetric in spin. Finally, 
fermionic antisymmetry of the Cooper pair wavefunction then
forces the Cooper pair to be antisymmetric in flavor.

Pairing between different colors/flavors can occur easily when
they all have the same chemical potentials and Fermi momenta.
This is the situation at very high density, where the strange quark
mass is negligible.
However, even at the very center
of a compact star the quark number chemical potential
$\mu$ cannot be much larger than 500 MeV, meaning
that the strange quark mass $M_s$ (which is density dependent, lying
somewhere between its vacuum current mass of about 100 MeV and constituent
mass of about 500 MeV) cannot be neglected.
Furthermore, bulk matter, as relevant for a compact star, must be in weak equilibrium
and must be electrically and color
neutral\cite{Iida:2000ha,Amore:2001uf,Alford:2002kj,Steiner:2002gx,Huang:2002zd,Neumann:2002jm}
(possibly via mixing of oppositely charged phases).
All these factors work to separate the Fermi momenta of the three different
flavors of quarks, and thus disfavor the cross-species BCS pairing
that characterizes the CFL phase.
If we imagine beginning at asymptotically high densities and reducing
the density, and suppose that CFL pairing is disrupted by the heaviness
of the strange quark before color superconducting quark matter is superseded
by baryonic matter, the CFL phase must be replaced by some
phase of quark matter in which there is less, and less symmetric, pairing.


In the next few subsections we give a quick overview of the
expected phases of real-world quark matter. We restrict our
discussion to zero temperature because the critical temperatures
for most of the phases that we discuss are expected to be
of order $10~\MeV$ or higher, and the core temperature
of a neutron star is believed to drop below this value
within minutes (if not seconds) of its creation in a supernova.

\subsection{Kaon condensation: the CFL-$K^0$ phase}
Bedaque and Sch\"afer\cite{Bedaque:2001je,Kryjevski:2004jw} showed
that when the stress is not too large (high density), it may simply
modify the CFL pairing pattern by inducing a flavor rotation of the
condensate which can be interpreted as a condensate of ``$K^0$''
mesons, i.e.~the neutral anti-strange Goldstone bosons associated with
the chiral symmetry breaking.  This is the ``CFL-K0'' phase, which
breaks isospin.  The $K^0$-condensate can easily be suppressed by
instanton effects,\cite{Schafer:2002ty} but if these are ignored then
the kaon condensation occurs for $M_s \gtrsim m^{1/3}\De^{2/3}$ for
light ($u$ and $d$) quarks of mass $m$.
This was demonstrated using an effective theory of the Goldstone bosons, but
with some effort can also be seen in an NJL 
calculation.\cite{Buballa:2004sx,Forbes:2004ww}

\subsection{The gapless CFL  phase}

The nature of the next significant transition has been
studied in NJL model calculations which ignore the
$K^0$-condensation in the CFL phase and which assume
spatial 
homogeneity.\cite{Alford:2003fq,Alford:2004hz,Alford:2004nf,Ruster:2004eg,Fukushima:2004zq,Abuki:2004zk,Ruster:2005jc,Blaschke:2005uj}
It has been found that
the phase structure depends on the
strength of the pairing. If the pairing is sufficiently strong
(so that $\De_{0}\sim 100~\MeV$ where
$\De_{0}$ is what the CFL gap would be at $\mu\sim 500~\MeV$
if $M_s$ were zero)
then the CFL phase survives all the way down to the transition
to nuclear matter. 
For a wide range of parameter values, however,
we find something
more interesting. We can make a
rough quantitative analysis by expanding in powers of
$M_s^2/\mu^2$ and $\De/\mu$, and
ignoring the fact that the effective strange
quark mass may be different in different phases.\cite{Alford:2002kj}
Such an analysis shows that, within a spatially homogeneous ansatz,
as we come down in density 
we find a transition at $\mu \approx \half M_s^2/\De_{0}$
from CFL to another phase, the gapless CFL phase (gCFL).\cite{Alford:2003fq,Alford:2004hz}
In this phase, quarks of all three colors and all three flavors
still form ordinary Cooper pairs, with each pair having zero total
momentum, but there are regions of momentum space in which certain
quarks do not succeed in pairing, and these regions are bounded by
momenta at which certain fermionic quasiparticles are gapless. This
variation on BCS pairing --- in which the same species of fermions
that pair feature gapless quasiparticles --- was first proposed for
two flavor quark matter\cite{Shovkovy:2003uu,Huang:2003xd} and in an atomic
physics context.\cite{Gubankova:2003uj}

For $M_s^2/\mu\gtrsim 2\De_0$, the CFL phase has higher free energy
than the gCFL phase.
This follows from the energetic balance between
the cost of keeping the Fermi surfaces together and the benefit
of the pairing that can then occur. 
The leading effect of $M_s$ is like a shift in the chemical potential
of the strange quarks, so the $bd$ and $gs$ quarks feel ``effective
chemical potentials''
$\mu_{bd}^{\rm eff} = \mu - \twothirds \mu_8$ and
$\mu_{gs}^{\rm eff} = \mu  + \third \mu_8 -\frac{M_s^2}{2\mu}$.
In the CFL phase $\mu_8=-M_s^2/(2\mu)$,\cite{Alford:2002kj,Steiner:2002gx}
so $\mu_{bd}^{\rm eff} - \mu_{gs}^{\rm eff} = M_s^2/\mu$.
The CFL phase will be stable as long as the
pairing makes it energetically favorable to maintain equality of the
$bd$ and $gs$ Fermi momenta, despite their differing chemical
potentials.\cite{Rajagopal:2000ff,Alford:2003fq}
It becomes unstable when
the energy gained from turning a 
$gs$ quark near the common Fermi momentum into a $bd$ quark 
(namely $M_s^2/\mu$) exceeds the cost
in lost pairing energy $2\De_0$. 
So the CFL phase is stable when
\beq
\frac{M_s^2}{\mu} < 2\De_{0}\ .
\label{CFLstable}
\eeq
Including the effects of $K^0$-condensation expands the range of $M_s^2/\mu$
in which the CFL phase is stable by a few tens of 
percent.\cite{Kryjevski:2004jw,Buballa:2004sx,Forbes:2004ww}
For larger $M_s^2/\mu$, the CFL phase is      
replaced by some new phase with unpaired $bd$ quarks,
which  cannot be neutral unpaired
or 2SC quark matter because the 
new phase and the CFL phase must have the same 
free energy at the critical $M_s^2/\mu = 2\De_{0}$.

\begin{figure}[t]
\bc
\includegraphics[width=0.6\hsize,angle=0]{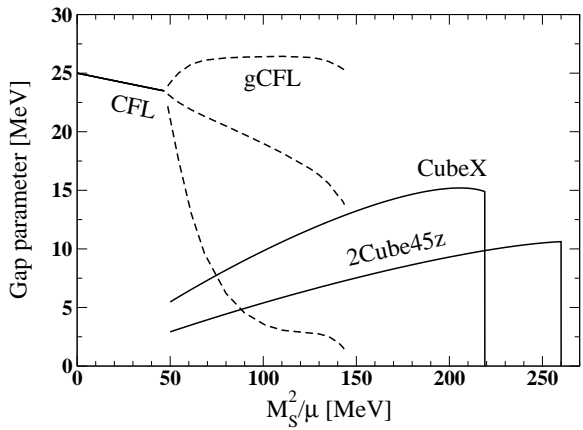}
\ec
\caption{Gap parameter $\Delta$ versus $M_s^2/\mu$ for the CFL and gCFL phases
as well as for crystalline color superconducting phases with two different
crystal structures described in Section \ref{sec:LOFFmultiple}.
These gap parameters have all been evaluated in a NJL model with
the interaction strength chosen such that the CFL gap parameter is 25 MeV
at $M_s^2/\mu=0$.  In the  gCFL phase, $\De_{ud} > \De_{us} > \De_{ds}$.
The calculation of the gap parameters in the crystalline phases has been
done in a Ginzburg-Landau approximation.
Recall that the splitting between Fermi surfaces is proportional 
to $M_s^2/\mu$, and that small (large)
$M_s^2/\mu$ corresponds to high (low) density.
}
\label{deltavsx}
\end{figure}

\begin{figure}[t]
\bc
\includegraphics[width=0.6\hsize,angle=0]{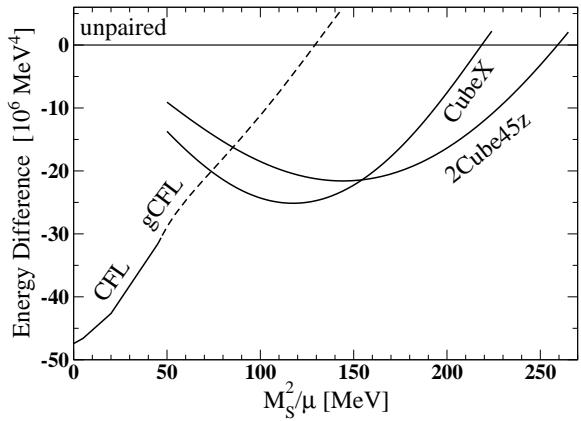}
\ec
\caption{
Free energy $\Omega$ relative to that of neutral unpaired quark matter
versus $M_s^2/\mu$ for the CFL, gCFL and crystalline phases whose gap
parameters are plotted in Fig.~\ref{deltavsx}.  Recall that the gCFL
phase is known to be unstable, meaning that in the regime where the
gCFL phase free energy is plotted, the true ground state of
three-flavor quark matter must be some phase whose free energy lies
below the dashed line. We see that the three-flavor crystalline color
superconducting quark matter phases with the most favorable crystal
structures that we have found, namely 2Cube45z and CubeX described in
(\ref{2Cube45zStructure}) and (\ref{CubeXStructure}), have
sufficiently robust condensation energy (sufficiently negative
$\Omega$) that they are candidates to be the ground state of matter
over a wide swath of $M_s^2/\mu$, meaning over a wide range of
densities.
}
\label{omegavsx}
\end{figure}

The obvious approach to finding this phase is to perform a NJL model
calculation with a general ansatz for the pairing that includes
differences between the flavors, for example by allowing different
pairing strengths $\De_{ud}$, $\De_{ds}$, $\De_{us}$. This was done in
Refs.~\refcite{Alford:2003fq,Alford:2004hz}, and the resultant
``gCFL'' phase was described in detail. In Figs.~\ref{deltavsx}
and~\ref{omegavsx} we show the results.  The gCFL phase takes over
from CFL at $M_s^2/\mu \approx 2 \De_{0}$, and remains favored beyond
the value $M_s^2/\mu \approx 4 \De_{0}$ at which the CFL phase would
become unfavored.

\subsection{Beyond gapless CFL}

The arguments above led us to the conclusion that the favored phase of
quark matter at the highest densities is the CFL phase, and that as
the density is decreased there is a transition to another color
superconducting phase, the gapless CFL phase.  However, it turns out
that the gCFL phase is itself unstable, and that in the regime where
the gCFL phase has a lower free energy than both the CFL phase and
unpaired quark matter, there must be some other phase with lower free
energy still.  The nature of that phase has to date not been
determined rigorously, and in subsequent subsections we shall discuss
various possibilities.  However, the crystalline color superconducting
phases analyzed very recently in Ref.~\refcite{Rajagopal:2006ig} whose
free energies are shown in Fig.~\ref{omegavsx} appear particularly
robust, offering the possibility that over a wide range of densities
they are the ground state of dense matter. We shall therefore
discuss these phases at much greater length in Section \ref{sec:LOFF}.

In all the contexts in which they have been investigated (whether in atomic
physics or the g2SC phase in two-flavor quark matter or the gCFL phase)
the gapless paired state turns out to suffer
from a ``magnetic instability'': it can lower its energy by the
formation of counter-propagating 
currents.\cite{Huang:2004bg,Casalbuoni:2004tb,Giannakis:2004pf,Alford:2005qw,Huang:2005pv,Fukushima:2005cm,Gorbar:2005rx,Kryjevski:2005qq,Gorbar:2006up,Iida:2006df,Fukushima:2006su,Gerhold:2006dt}
In the atomic
physics context, the resolution of the instability is phase
separation into macroscopic regions of two phases in one of which
standard BCS pairing occurs and in the other of which no pairing
occurs.\cite{Bedaque:2003hi,KetterleImbalancedSpin,HuletPhaseSeparation}
Phase separation into electrically charged but color neutral
phases is also a possibility in two-flavor quark matter.\cite{Reddy:2004my}
In three-flavor quark matter, where the instability of the gCFL
phase has been established in Refs.~\refcite{Casalbuoni:2004tb,Fukushima:2005cm}, phase
coexistence would require coexisting components with opposite color
charges, in addition to opposite electric charges, making it very
unlikely that a phase separated solution can have lower energy than
the gCFL phase.\cite{Alford:2004hz,Alford:2004nf}   Furthermore,
color superconducting phases which are less symmetric than the CFL
phase but still involve only conventional BCS pairing, for example
the much-studied 2SC phase in which only two colors of up and down
quarks pair~\cite{Bailin:1983bm,Alford:1997zt,Rapp:1997zu} but
including also many other possibilities,\cite{Rajagopal:2005dg}
cannot be the resolution of the gCFL
instability.\cite{Alford:2002kj,Rajagopal:2005dg} It seems likely,
therefore, that a ground state with counter-propagating currents is
required.  This could take the form of a crystalline color
superconductor\cite{Alford:2000ze,Bowers:2001ip,Casalbuoni:2001gt,Leibovich:2001xr,Kundu:2001tt,Bowers:2002xr,Casalbuoni:2003wh,Casalbuoni:2003sa,Casalbuoni:2004wm,Casalbuoni:2005zp,Ciminale:2006sm,Mannarelli:2006fy,Rajagopal:2006ig}
--- the QCD analogue of a form of non-BCS pairing first considered
by Larkin, Ovchinnikov, Fulde and Ferrell.\cite{LOFF} Or, given
that the CFL phase itself is likely augmented by kaon
condensation,\cite{Bedaque:2001je,Kryjevski:2004jw,Buballa:2004sx,Forbes:2004ww} it could take
the form of a phase in which a CFL kaon condensate carries a current
in one direction balanced by a counter-propagating current in the
opposite direction carried by gapless quark
quasiparticles.\cite{Kryjevski:2005qq,Gerhold:2006dt}

\begin{figure}[t]
 \bc
 \includegraphics[width=0.6\hsize]{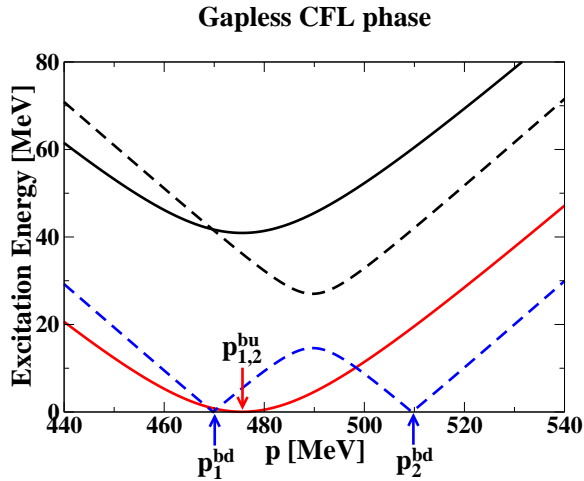}
 \ec
\caption{
Dispersion relations of the lightest quasiquark excitations
in the gCFL phase, at $\mu=500~\MeV$,
with $M_s=200~\MeV$ and  interaction strength
such that the CFL gap parameter at $M_s=0$ would be $\De_{0}=25~\MeV$.
Note that in there is a gapless mode with a 
{\em quadratic}\/ dispersion relation
(energy reaching zero at momentum $p^{bu}_{1,2}$) as well
as two gapless modes with more conventional linear dispersion
relations. 
}
\label{fig:disprel}
\end{figure}

The instability of the gCFL phase appears to be related to one of its
most interesting features, namely
the presence of gapless fermionic excitations around the ground state.
These are illustrated in Fig.~\ref{fig:disprel}, which shows that
there is one mode (the $bu$-$rs$ quasiparticle) with
an unusual quadratic dispersion relation, which
is expected to give rise to a parametrically
enhanced heat capacity and neutrino emissivity  and 
anomalous transport properties.\cite{Alford:2004zr}
The instability manifests itself
in imaginary Meissner masses $M_M$ for some of the gluons.
$M_M^2$ is the
low-momentum current-current two-point function, 
and $M_M^2/(g^2\De^2)$, with $g$ the gauge coupling, is the 
coefficient of the 
gradient term in the effective theory of small fluctuations around the
ground-state condensate.
The fact that we find a negative value when the quasiparticles are gapless
indicates an instability 
towards spontaneous breaking of translational invariance.
Calculations in a simple two-species model\cite{Alford:2005qw}
show that imaginary $M_M$ is generically associated with
the presence of gapless charged fermionic modes.


The earliest calculations for the three-flavor case show that even a
very simple ansatz for the crystal structure yields a crystalline
color superconducting state that has lower free energy than gCFL in
the region where the gCFL$\to$unpaired transition
occurs.\cite{Casalbuoni:2005zp,Mannarelli:2006fy} It is reasonable,
based on what was found in the two-flavor case in
Ref.~\refcite{Bowers:2002xr}, to expect that when the full space of
crystal structures is explored, the crystalline color superconducting
state will be preferred to gCFL over a much wider range of the stress
parameter $M_s^2/(\Delta_0\mu)$, and this expectation has recently
been confirmed by explicit calculation.\cite{Rajagopal:2006ig} It is
conceivable that the whole gCFL region is actually a crystalline
region, but the results shown in Fig.~\ref{omegavsx} that we describe
in Section \ref{sec:LOFF} suggest that there is still room for other
possibilities (like the current-carrying meson condensate) at the
highest densities in the gCFL regime.

An alternative explanation of the consequences of the gCFL instability
was advanced by Hong\cite{Hong:2005jy}
(see also Ref.~\refcite{Huang:2003xd}): since the
instability is generically associated with the presence of gapless fermionic
modes, and the BCS mechanism implies that any gapless fermionic mode
is unstable to Cooper pairing in the most attractive channel, one might
expect that the instability will simply be resolved by ``secondary pairing''.
This means
the formation of a $\<qq\>$ condensate where $q$ is either one of the gapless
quasiparticles whose dispersion relation is shown in Fig.~\ref{fig:disprel}.
After the formation of such a secondary condensate, the linear 
gapless dispersion relations
would be modified by ``rounding out'' of the corner where the energy
falls to zero, leaving a secondary energy gap $\De_s$, which
renders the mode gapped, and removes the instability.
In the case of the quadratically gapless mode there is a greatly
increased density of states at low energy 
(in fact, the density of states diverges as $E^{-1/2}$), so Hong
calculated that the secondary pairing should
be much stronger than would be predicted by BCS theory, 
and he specifically predicted $\Delta_s \propto G_s^2$ 
for coupling strength $G_s$, as compared with
the standard BCS result
$\Delta \propto \exp(-{\rm const}/G_s)$.

This possibility was worked out in a two-species NJL model in
Ref.~\refcite{Alford:2005kj}. This allowed a
detailed exploration of the strength of secondary pairing. The
calculation confirmed Hong's prediction that in typical secondary
channels $\Delta_s \propto G_s^2$. However, in all the secondary
channels that were analyzed it was found that the secondary gap, even
with this enhancement, is from ten to hundreds of times smaller than
the primary gap at reasonable values of the secondary coupling. 
This shows that that secondary pairing
does not generically resolve the magnetic
instability of the gapless phase, since it indicates that there is
a temperature range $\De_s \ll T \ll \De_p$ in which there is primary
pairing (of strength $\De_p$)
but no secondary pairing, and at those temperatures the
instability problem would arise again.

\subsection{Crystalline pairing}
The pairing patterns discussed so far have been
translationally invariant. But in the region of parameter space
where cross-species pairing is excluded by stresses
that pull apart the Fermi surfaces, one expects a position-dependent
pairing known as the crystalline color superconducting phase.\cite{LOFF,Alford:2000ze,Bowers:2001ip,Casalbuoni:2001gt,Leibovich:2001xr,Kundu:2001tt,Bowers:2002xr,Casalbuoni:2003wh,Casalbuoni:2003sa,Casalbuoni:2004wm,Casalbuoni:2005zp,Ciminale:2006sm,Mannarelli:2006fy,Rajagopal:2006ig}
This arises because one way
to achieve pairing between different flavors while accomodating the
tendency for the Fermi momenta to separate is to 
allow Cooper pairs with nonzero total momentum yielding
pairing over parts
of the Fermi surfaces.
We describe this phase extensively in Section \ref{sec:LOFF}.
As we will discuss there, this phase may resolve the gCFL phase's
stability problems.  

\subsection{Single-flavor pairing}
At densities that are so low that $M_s$ puts such a significant
stress on the pairing pattern that even the crystalline phase becomes
unfavored, we expect  a transition to 
a phase with no cross-species
pairing at all. (This regime will only arise if $\Delta_0$ is so small that
very large values of $M_s^2/(\mu\Delta_0)$ can arise 
without $\mu$ being taken so small
that nuclear matter becomes favored.)

\begin{figure}[t]
\includegraphics[scale=0.25]{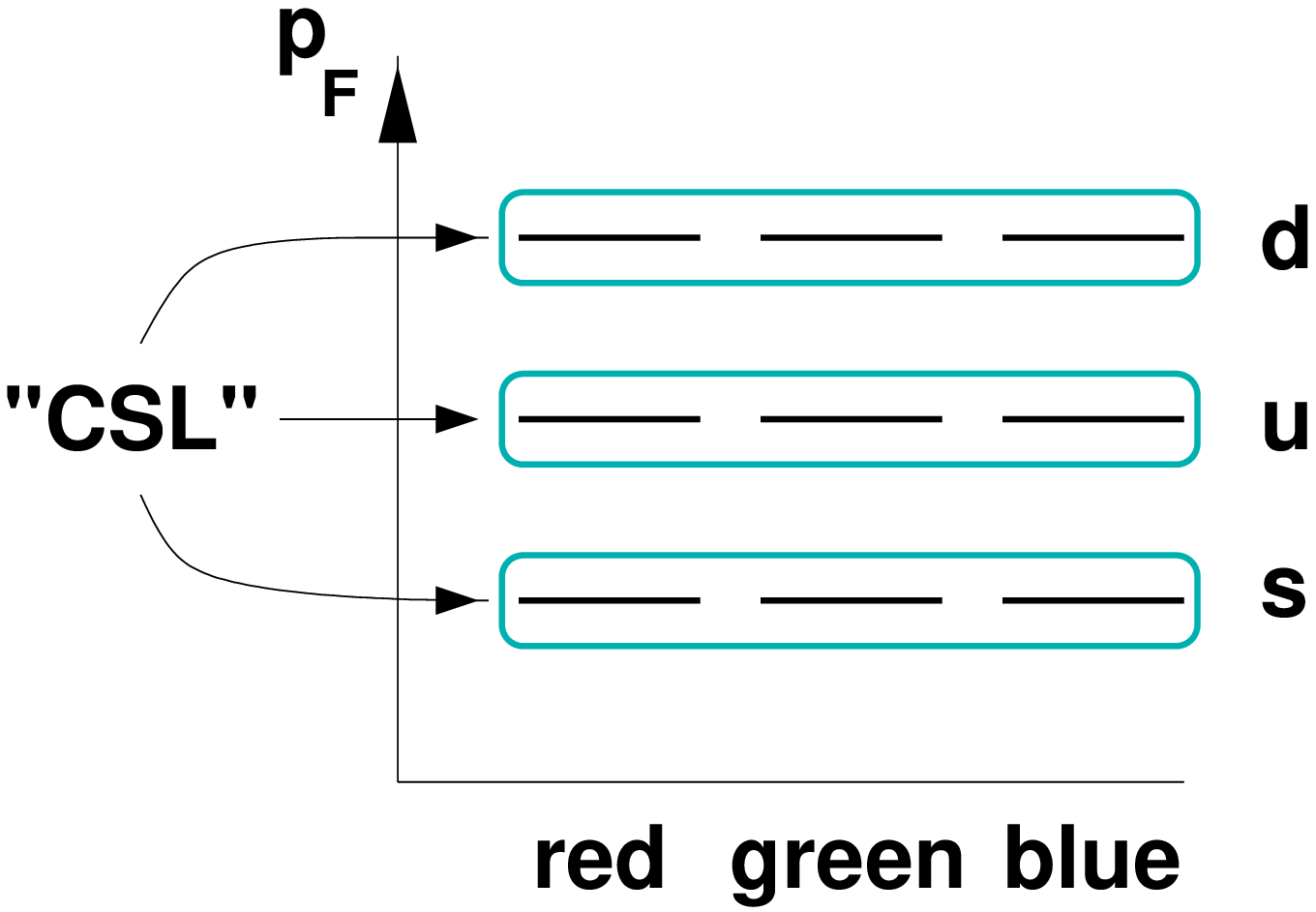}
\hspace{1em}
\includegraphics[scale=0.25]{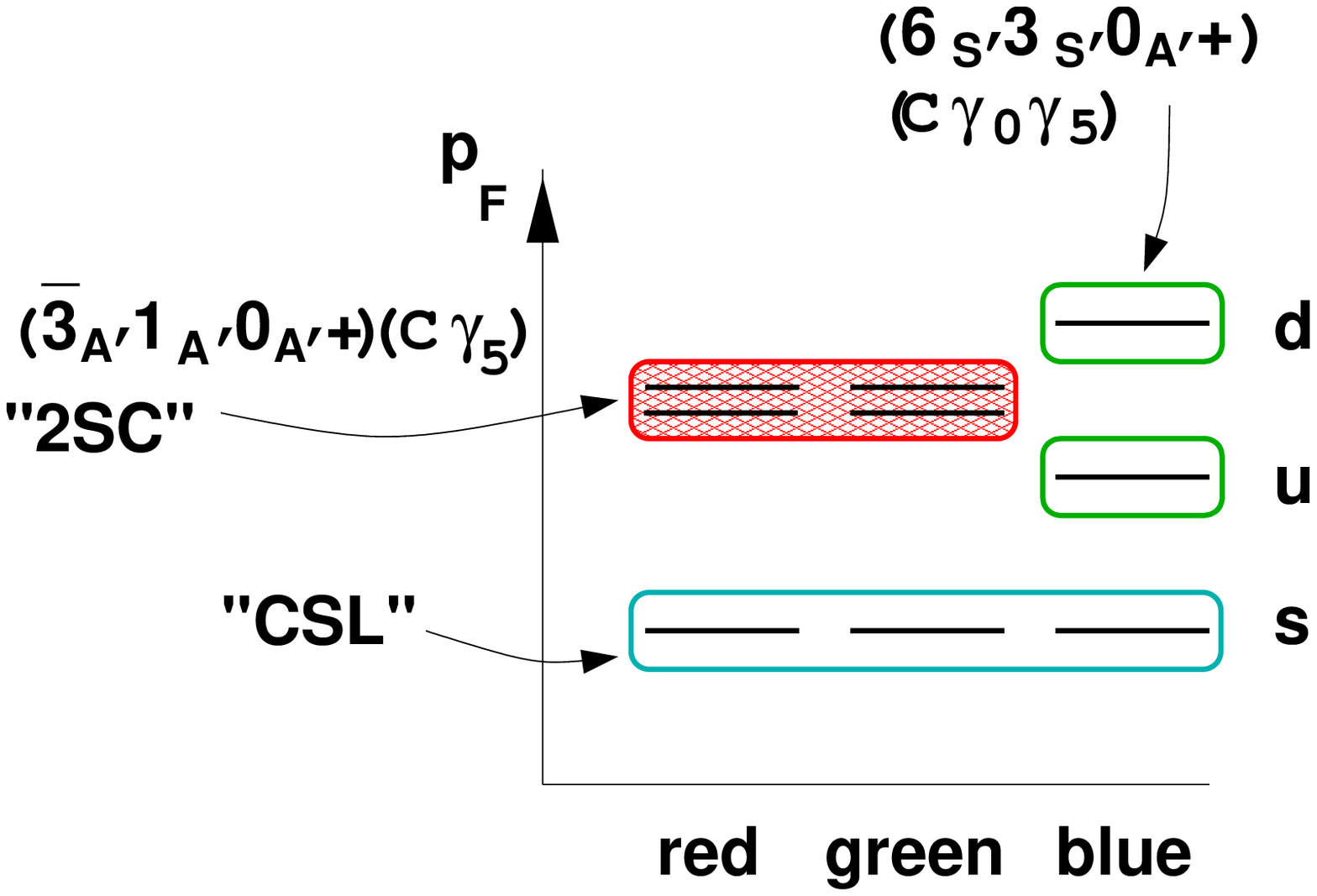}
\hspace{1em}
\includegraphics[scale=0.25]{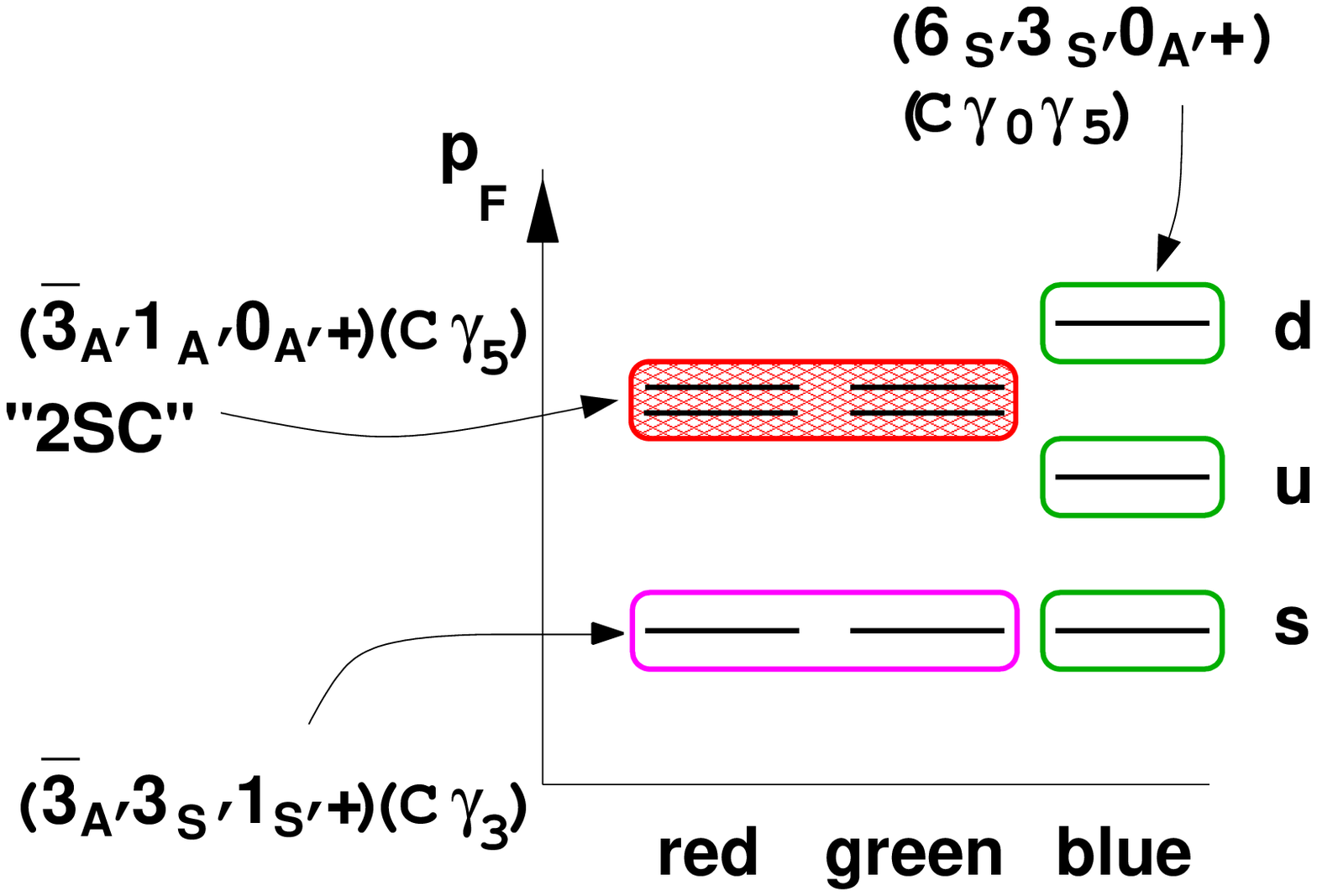}
\caption{Pictorial representation of CSL pairing, 2SC+CSL pairing,
and 2SC+1SC pairing in neutral quark matter. 
See Ref.~\protect\refcite{Alford:2005yy}.
}
\label{fig:Fermi_momenta}
\end{figure}

\noindent\underline{``Unpaired'' quark matter}.
In most NJL studies, matter with no cross-species pairing at all is
described as
as ``unpaired'' quark matter. However, it is well known that
there are attractive channels for a single flavor pairing, although they
are much weaker than the 2SC and CFL 
channels.\cite{Schafer:2000tw,Buballa:2002wy,Alford:2002rz,Schmitt:2002sc,Schmitt:2004et}
Calculations using NJL models and single-gluon exchange agree
that the favored phase in this case is the color-spin-locked (CSL) 
phase,\cite{Schafer:2000tw}
in which all three colors of each flavor, with each pair of colors
correlated with a particular direction for the spin.
This phase does not break rotational symmetry.
See first panel of Fig.~\ref{fig:Fermi_momenta}.

\noindent\underline{``Unpaired'' species in 2SC quark matter}.

If the stress is large enough to destroy CFL pairing, and furthermore
is large enough to preclude cross-species pairing in which the Cooper
pairs have nonzero momentum and form a crystalline color
superconducting phase, we can ask whether there is any pattern of BCS
pairing which can persist in neutral quark matter with arbitrarily
large splitting between Fermi surfaces.  The answer to this is no, as
demonstrated via an exhaustive search in
Ref.~\refcite{Rajagopal:2005dg}.  Although unlikely, it remains a
possibility that up and down quarks with two colors could pair in the
2SC pattern. This does not happen in NJL models in which $M_s$ is a
parameter.  However, in models in which one instead solves
self-consistently for the quark masses, the 2SC phase tends to become
more robust.  It remains to be seen whether this relative
strengthening of the 2SC phase in such models is sufficient to allow
it to exist at values of $M_s^2/\mu$ which are larger than those where
the crystalline phases shown in Fig.~\ref{omegavsx} and discussed at
length in Section \ref{sec:LOFF} are favored.

If there is a regime in which the 2SC phase survives,
this leaves the blue quarks unpaired.
In that case one might expect a ``2SC+CSL'' pattern, illustrated
in the second panel of Fig.~\ref{fig:Fermi_momenta},
which would again be rotationally symmetric. However,
the 2SC pattern breaks the color symmetry, and
in order to maintain color neutrality, a color chemical potential
is generated, which also affects the unpaired strange quarks,
splitting the Fermi momentum of the blue strange quarks
away from that of the red and green strange quarks.
This is a small effect, but so is the CSL pairing gap, 
and NJL model calculations indicate
that the color chemical potential typically destroys CSL 
pairing of the strange quarks.\cite{Alford:2005yy}
The system falls back on the next best alternative, which
is spin-1 pairing of the red and green strange quarks
(third panel of Fig.~\ref{fig:Fermi_momenta}).

Single-flavor pairing phases have much lower critical temperatures 
than multi-flavor phases like the CFL or crystalline
phases, perhaps as large as a few MeV, more typically
in the eV to many keV range
\cite{Schafer:2000tw,Buballa:2002wy,Alford:2002rz,Schmitt:2002sc,Schmitt:2004et},
so they are expected to play a role late in the life of a neutron star.

\subsection{Mixed Phases}
Another way for a system to deal with a stress on its
pairing pattern is phase separation. In the context of quark
matter this corresponds to relaxing the requirement of local charge
neutrality, and requiring neutrality only over long distances,
so we allow a mixture of a positively charged and a negatively charged
phase, with a common pressure and a common value of the electron
chemical potential $\mu_e$ that is not equal to the neutrality value
for either phase. Such a mixture of nuclear and CFL quark matter
was studied in Ref.~\refcite{Alford:2001zr}. In three-flavor quark matter
it has been found that as long as we require local color neutrality 
such mixed phases are not the favored response to the stress imposed
by the strange quark mass.\cite{Alford:2003fq,Alford:2004hz,Alford:2004nf}
 Phases involving
color charge separation have been studied\cite{Neumann:2002jm} but it seems
likely that the energy cost of the color-electric fields will disfavor them.

\section{The crystallography of three-flavor quark matter}
\label{sec:LOFF}

In this section we review recent work by Rajagopal and
Sharma.\cite{Rajagopal:2006ig} Here, we focus on explaining the
context, the model, the results and the implications. We do not
describe the calculations.

\subsection{Introduction and context}
\label{sec:LOFFintro}

The investigation of crystalline color superconductivity in
three-flavor QCD was initiated in Ref.~\refcite{Casalbuoni:2005zp}.
Although such phases seem to be free from magnetic
instability,\cite{Ciminale:2006sm} it remains to be determined whether
such a phase can have a lower free energy than the meson current
phase, making it a possible resolution to the gCFL instability.  The
simplest ``crystal'' structures do not 
suffice,\cite{Casalbuoni:2005zp,Mannarelli:2006fy} but experience in
the two-flavor context\cite{Bowers:2002xr} suggests that realistic
crystal structures constructed from more plane waves will prove to
be qualitatively more robust.   The results of Ref.~\refcite{Rajagopal:2006ig} 
confirm this expectation.

Determining the favored crystal structure(s) in the
crystalline color superconducting phase(s) of three-flavor QCD requires
determining the gaps and comparing the free energies for very many candidate structures,
as there are even more possibilities than the many that were investigated in
the two-flavor context.\cite{Bowers:2002xr}   As there, we shall make
a Ginzburg-Landau approximation.
This approximation is controlled if $\Delta \ll \Delta_0$,
where $\Delta$ is the gap parameter of the crystalline color superconducting
phase itself and $\Delta_0$ is the gap parameter in the CFL phase that would
occur if $M_s$ were zero.    We shall find that the most favored crystal
structures can have $\Delta/\Delta_0$ as large as $\sim 1/2$, meaning
that we are pushing the approximation hard and so should not trust it quantitatively.
Earlier work on
a particularly simple one parameter family of ``crystal'' structures in
three-flavor quark matter,\cite{Mannarelli:2006fy} simple enough that the analysis could be done
both with and without the Ginzburg-Landau approximation, demonstrates that
the approximation works when it should and that, at least
for very simple crystal structures,
when it breaks down it always underestimates the gap $\Delta$ and the
condensation energy.  Furthermore, the Ginzburg-Landau approximation
correctly determines which crystal structure among the one parameter family
that we analyzed in Ref.~\refcite{Mannarelli:2006fy}  
has the largest gap and lowest free energy.  

The underlying microscopic theory that we use is
an NJL
model in which the QCD interaction between
quarks is replaced by a point-like four-quark interaction with the quantum
numbers of single-gluon exchange, analyzed in mean field theory.
This is not a controlled approximation.
However, it suffices for our purposes: because this model has  attraction
in the same channels as in QCD, its high density phase is the CFL phase; and, the
Fermi surface splitting effects whose
qualitative consequences we wish to study can all be built
into the model.  Note that we
shall assume throughout that $\Delta_0\ll \mu$.  This weak coupling assumption
means that the pairing is dominated by modes near the Fermi surfaces.
Quantitatively,  this means that results for the gaps and condensation energies
of candidate crystalline phases are independent of the cutoff in the NJL model
when expressed in terms of the CFL gap $\Delta_0$: if the cutoff is changed with
the NJL coupling constant adjusted so that $\Delta_0$ stays fixed, the gaps and
condensation energies for the candidate crystalline phases also stay fixed.
This makes the NJL model valuable for making the comparisons
that are our goal.

We shall consider crystal structures in which there are two condensates
\begin{eqnarray}
\langle ud \rangle &\sim& \Delta_3 \sum_{a} \exp\left(2i\q{3}{a} \cdot {\bf r}\right)\nonumber\\
\langle us \rangle &\sim& \Delta_2 \sum_{a} \exp\left(2i\q{2}{a}\cdot {\bf r} \right)\ .
\label{udanduscondensates}
\end{eqnarray}
As in Refs.~\refcite{Casalbuoni:2005zp,Mannarelli:2006fy}, 
we neglect $\langle ds \rangle$ pairing because the $d$ and $s$
Fermi surfaces  are twice as far apart from each other as each is from
the intervening $u$ Fermi surface.  Were we to set $\Delta_2$ to zero, treating
only $\langle ud \rangle$ pairing, we would recover the two-flavor Ginzburg-Landau
analysis of Ref.~\refcite{Bowers:2002xr}.   There, it was found that the best
choice of crystal structure was one in which pairing occurs for a set
of eight ${\bf q}_3^a$'s pointing at the corners of a cube in momentum space,
yielding a condensate with face-centered cubic symmetry.  The 
analyses of three-flavor crystalline color superconductivity in 
Refs.~\refcite{Casalbuoni:2005zp,Mannarelli:2006fy} introduced nonzero $\Delta_2$,
but made the simplifying ansatz that pairing occurs only for a single ${\bf q}_3$
and a single ${\bf q}_2$. We consider crystal structures with up to eight
${\bf q}_3^a$'s and up to eight ${\bf q}_2^a$'s.  

We evaluate the
free energy $\Omega(\Delta_2,\Delta_3)$ for each three-flavor crystal structure in
a Ginzburg-Landau expansion in powers of the $\Delta$'s. We work
to order $\Delta_2^p\Delta_3^q$ with $p+q=6$.  At sextic order, 
we find that $\Omega(\Delta,\Delta)$ is positive for large $\Delta$
for all the crystal structures that we investigate.\cite{Rajagopal:2006ig}  This is in marked
contrast to the result that
many two-flavor crystal structures have negative sextic terms,
with free energies that are unbounded from below when the 
Ginzburg-Landau expansion is stopped at sextic order.\cite{Bowers:2002xr}  Because we 
find positive sextic terms, we are able to use our sextic Ginzburg-Landau expansion
to evaluate $\Delta$ and $\Omega(\Delta,\Delta)$ for all the structures that we analyze.

We can then identify the two most favorable crystal structures,
2Cube45z and CubeX, which are described further in
Section \ref{sec:LOFFmultiple}.
To a large degree, our argument that these two structures are the most
favorable relies only on two qualitative factors, which we explain
in Section \ref{sec:LOFF2plane}. CubeX and 2Cube45z have gap parameters that are large:
a third to a half of $\Delta_0$. We have already
discussed the implications of this for the Ginzburg-Landau approximation
in Section \ref{sec:LOFFintro} above.
The large gaps naturally lead to
large condensation energies, easily 1/3 to 1/2 of that in the CFL phase
with $M_s=0$, which is  $3\Delta_0^2\mu^2/\pi^2$.  This is remarkable, given
the only quarks that pair are those lying on (admittedly many) rings on
the Fermi surfaces, whereas in the CFL phase with $M_s=0$ pairing
occurs over the entire $u$, $d$ and $s$ Fermi surfaces.  
The gapless CFL (gCFL)
phase provides a useful
comparison at nonzero $M_s$. For $2 \Delta_0 < M_s^2/\mu < 5.2 \Delta_0$,
model analyses that are restricted to isotropic phases
predict a gCFL phase,\cite{Alford:2003fq,Alford:2004hz,Fukushima:2004zq} finding
this phase to have lower free energy than either the CFL phase or unpaired quark 
matter, as shown in Fig.~\ref{omegavsx}.
However, this phase
is unstable to the formation of current-carrying
condensates.\cite{Huang:2004bg,Casalbuoni:2004tb,Giannakis:2004pf,Alford:2005qw,Huang:2005pv,Fukushima:2005cm,Gorbar:2005rx,Kryjevski:2005qq,Gorbar:2006up,Iida:2006df,Fukushima:2006su,Gerhold:2006dt}
and so it cannot be the ground state.  The true ground state must have lower
free energy than that of the gCFL phase, 
and for this reason the gCFL free energy provides a useful benchmark.
We find that three-flavor crystalline color superconducting quark matter 
with one or other of the two crystal structures that we argue
are most favorable has lower free energy (greater condensation energy)
than CFL quark matter, gCFL quark matter, and unpaired quark matter
for
\beq
2.9 \Delta_0 < \frac{M_s^2}{\mu} < 10.4 \Delta_0\ .
\label{WideLOFFWindow}
\eeq
(See Fig.~\ref{omegavsx} above.)
This window in parameter space is in no sense narrow.
Our results therefore indicate that three-flavor crystalline quark matter
will occur over a wide range of densities. (That is unless the pairing between
quarks is so strong, i.e. $\Delta_0$ so large making $M_s^2/\Delta_0$ so
small, that quark matter is in the 
CFL phase all the way down to the density at which quark matter
is superseded by nuclear matter.)
However, our results  also indicate
that unless the Ginzburg-Landau approximation is underestimating
the condensation energy of the
crystalline phase by about a factor of two, there is a fraction
of the ``gCFL window'' (with $2 \Delta_0 < M_s^2/\mu < 2.9 \Delta_0$, in
the Ginzburg-Landau approximation) in which no crystalline phase
has lower free energy than the gCFL phase.  This is thus the most
likely regime in which to find the current-carrying meson condensates
of Refs.~\refcite{Kryjevski:2005qq,Gerhold:2006dt}.

\subsection{Model, simplifications and ansatz}

\subsubsection{Neutral unpaired three-flavor quark matter}

The analysis of beta-equilibrated neutral three-flavor quark matter 
in an NJL model has been
discussed in detail elsewhere (For example, Ref.~\refcite{Alford:2004hz}), and 
here we just summarize the main points.
We assume that the up and down quarks are massless. The strange
quark mass is a parameter $M_s$. We couple a chemical potential 
$\mu$ to quark number, and always work at $\mu=500$~MeV.
To ensure neutrality under color and electromagnetism we couple 
a chemical (electrostatic) potential $\mu_e$ to negative electric charge,
and chemical (color-electrostatic) potentials $\mu_3$ and $\mu_8$
to the diagonal generators of $SU(3)_{\rm color}$. In full QCD
these gauged chemical potentials would be the zeroth components of the 
corresponding gauge fields, and would naturally be forced to their 
neutrality values.\cite{Alford:2002kj,Gerhold:2003js} In the NJL context,
we enforce neutrality explicitly by choosing them to satisfy the
neutrality conditions
$\frac{\partial \Omega}{\partial\mu_e} =
\frac{\partial \Omega}{\partial\mu_3} =
\frac{\partial \Omega}{\partial\mu_8} = 0$. 
The assumption of beta-equilibrium is built into the calculation via
the fact that the only flavor-dependent chemical potential we
introduce is $\mu_e$, ensuring for example that the chemical
potentials of $d$ and $s$ quarks with the same color must be equal.

The presence of a nonzero strange quark mass, combined with the
requirements of equilibrium under weak interactions and
charge neutrality, drives down the number of strange quarks,
and introduces electrons and additional down quarks. We assume that this
is a small effect, i.e.~we expand in powers of
$\mu_e/\mu$ and $M_s^2/\mu^2$. Working to linear order, 
we can simply treat the strange quark mass as 
a shift of $-M_s^2/(2\mu)$ in the strange quark chemical potential.
(The corrections from higher-order terms in $M_s^2/\mu^2$
in an NJL analysis of a two-flavor crystalline
color superconductor have been evaluated and found
to be small,\cite{Kundu:2001tt} and we expect the same to be true here.)
In this approximation we find
the Fermi momenta of the quarks and electrons are given by
\begin{eqnarray}
p_F^d &=& \mu+\frac{\mu_e}{3}\qquad p_F^u = \mu-\frac{2 \mu_e}{3}\qquad p_F^e = \mu_e
\nonumber\\
p_F^s &=& \sqrt{\left(\mu+\frac{\mu_e}{3}\right)^2- M_s^2} \approx \mu + \frac{\mu_e}{3}
-\frac{M_s^2}{2\mu}
\label{pF1}
\end{eqnarray}
and their free energy is
\bea
\Omega_{\rm unpaired} &=& -\frac{ 
3 \left( p_F^u\right)^4 + 3 \left(p_F^d\right)^4 + \left(p_F^e\right)^4 }{12\pi^2}
+ \frac{3}{\pi^2}\int_0^{p_F^s}p^2 dp \left(\sqrt{p^2+M_s^2}-\mu-\frac{\mu_e}{3}\right)
\nonumber\\
&\approx & -\frac{3}{4\pi^2}\left(\mu^4 -\mu^2 M_s^2\right)+ \frac{1}{2\pi^2}\mu M_s^2 \mu_e -
\frac{1}{\pi^2}\mu^2\mu_e^2+\ldots
\label{OmegaUnpaired}
\eea
To this order, electric neutrality requires
\beq
\mu_e=\frac{M_s^2}{4\mu}\ ,
\label{mueneutral}
\eeq
and because we are working to lowest order in $M_s^2/\mu^2$
we need no longer be careful about the
distinction between $p_F$'s and $\mu$'s, as we can simply think of the three
flavors of quarks as if they have chemical potentials
\begin{eqnarray}
\mu_d &=& \mu_u + 2 \delta\mu_3 \nonumber\\
\mu_u &=&p_F^u \nonumber\\
\mu_s &=& \mu_u - 2 \delta\mu_2
\label{pF3}
\end{eqnarray}
where, using (\ref{mueneutral}), $p_F^u = \mu-\frac{M_s^2}{6\mu}$ and
\beq
\delta\mu_3 = \delta\mu_2 = \frac{M_s^2}{8\mu}\equiv \delta\mu \ ,
\eeq
with the choice of subscripts indicating that
$2\delta\mu_2$ is the
splitting between the Fermi surfaces for quarks 1 and 3 and
$2\delta\mu_3$ is that between the Fermi surfaces for quarks 1 and 2,
identifying $u,d,s$ with $1,2,3$.

\subsubsection{BCS pairing and neutrality}

BCS pairing introduces qualitative changes into the analysis of 
neutrality.\cite{Rajagopal:2000ff,Alford:2002kj,Steiner:2002gx,Alford:2003fq}  For example,
in the CFL phase $\mu_e=0$ and $\mu_8$ is nonzero and of order $M_s^2/\mu$.
This arises because the construction of a phase in which BCS pairing occurs between
fermions whose Fermi surface would be split in the absence of pairing can be
described as follows. 
First, adjust the Fermi surfaces of those fermions that pair to make
them equal. This costs a free energy price of order $\delta\mu^2\mu^2$.  And, it
changes the relation between the chemical potentials and the particle numbers,
meaning that the $\mu$'s required for neutrality can change qualitatively as
happens in the CFL example. Second, pair.
This yields a free energy benefit of order $\Delta_0^2\mu^2$, where $\Delta_0$
is the gap parameter describing the BCS pairing.
Hence, BCS pairing will only
occur if the attraction between  the fermions is large enough that
$\Delta_0 \gtrsim \delta\mu$.  In the CFL context, in which $\langle ud \rangle$,
$\langle us \rangle$ and $\langle ds \rangle$ pairing is fighting against the
splitting between the $d$, $u$ and $s$ Fermi surfaces described above, it
turns out that CFL pairing can occur if
$\Delta_0>4\delta\mu=M_s^2/(2\mu)$,\cite{Alford:2003fq}
a criterion that is
reduced somewhat by kaon condensation which
acts to stabilize CFL pairing.\cite{Kryjevski:2004jw,Forbes:2004ww}

We are now considering quark matter at densities that are
low enough ($\mu<M_s^2/(2\Delta_0)$) that CFL pairing is not possible.
The gap parameter $\Delta_0$ that would characterize the CFL
phase if $M_s^2$ and $\delta\mu$ were zero
is nevertheless an important scale in our problem, as it
quantifies the strength of the attraction between quarks.
Estimates of the magnitude of $\Delta_0$ are typically in the tens of MeV,
perhaps as large as 100 MeV.\cite{Reviews}  We shall treat $\Delta_0$ as a parameter,
and quote results for $\Delta_0=25$~MeV, although 
our results can easily be scaled to any value of $\Delta_0$ 
as long as the weak-coupling approximation $\Delta_0\ll\mu$ is 
respected.\cite{Rajagopal:2006ig}

\subsubsection{Crystalline color superconductivity in two-flavor quark matter}
\label{sec:LOFF2flav}

Crystalline color superconductivity can be thought of as the answer
to the question: ``Is there a way to pair quarks at differing Fermi
surfaces without first equalizing their Fermi momenta, given that
doing so exacts a cost?" The answer is ``Yes, but it requires Cooper
pairs with nonzero total momentum."  Ordinary BCS pairing pairs
quarks with momenta ${\bf p}$ and $-{\bf p}$, meaning that if the
Fermi surfaces are split at most one member of a pair can be at its
Fermi surface.  In the crystalline color superconducting phase,
pairs with total momentum $2\bf q$ condense, meaning that one member
of the pair has momentum ${\bf p}+{\bf q}$ and the other has
momentum $-{\bf p}+{\bf q}$ for some $\bf
p$.\cite{LOFF,Alford:2000ze}  Suppose for a moment that only $u$
and $d$ quarks pair, making the analyses of a two-flavor model found
in
Refs.~\refcite{Alford:2000ze,Bowers:2001ip,Casalbuoni:2001gt,Leibovich:2001xr,Kundu:2001tt,Bowers:2002xr,Casalbuoni:2003wh,Casalbuoni:2003sa,Casalbuoni:2004wm}
(and really going back to Ref.~\refcite{LOFF}) valid. We sketch the
results of this analysis in this subsection.

The simplest ``crystalline'' phase
is one in which only pairs with a single $\bf q$ condense, yielding a condensate
\beq
\langle \psi_u(x) C \gamma_5 \psi_d(x) \rangle \propto \Delta \exp(2i {\bf q}\cdot {\bf r})
\label{singleplanewave}
\eeq
that is modulated in space like a plane wave.   (Here and throughout, we shall denote
by ${\bf r}$ the spatial three-vector corresponding to the Lorentz four-vector $x$.)
Assuming that $\Delta\ll \delta\mu \ll \mu$,
the energetically favored value of $|{\bf q}|\equiv q$ turns out to be $q=\eta \delta\mu$, where
the proportionality constant $\eta$ is given by $\eta=1.1997$.\cite{LOFF,Alford:2000ze}
If $\eta$ were 1, then the only choice of ${\bf p}$ for which a Cooper pair
with momentum $(-\bf p+{\bf q},{\bf p}+{\bf q})$ would describe two quarks each on
their respective Fermi surfaces would correspond to a quark on
the north pole of one Fermi surface and a quark on the south pole of the other.
Instead, with $\eta>1$, the quarks on each Fermi surface that can pair lie
on one ring on each Fermi surface, the rings having opening angle
$2\cos^{-1}(1/\eta)=67.1^\circ$.  The energetic calculation that determines $\eta$
can be thought of as balancing the gain in pairing energy as $\eta$ is increased
beyond $1$, allowing quarks on larger rings to pair, against the kinetic energy cost
of Cooper pairs with greater total momentum.   If the $\Delta/\delta\mu\rightarrow 0$
Ginzburg-Landau limit is not assumed, the pairing rings change from circular lines
on the Fermi surfaces into
ribbons of thickness $\sim\Delta$ and angular extent $\sim \Delta/\delta\mu$.

After solving a gap equation for $\Delta$
and then evaluating the free energy of the phase with condensate (\ref{singleplanewave}),
one finds that this simplest ``crystalline'' phase is favored over two-flavor quark matter
with  either no pairing or BCS pairing only within  a narrow window
\beq
0.707\, \Delta_{\rm 2SC} <  \delta\mu  < 0.754\,\Delta_{\rm 2SC}\ ,
\label{LOFFwindow}
\eeq
where $\Delta_{\rm 2SC}$ is the gap parameter for the two-flavor phase 
with 2SC (2-flavor, 2-color)
BCS pairing  found at $\delta\mu=0$.
At the upper boundary of this window, $\Delta\rightarrow 0$ and one finds a second
order phase transition between the crystalline and unpaired phases.  At the lower boundary,
there is a first order transition between the crystalline and BCS paired phases.
The crystalline phase persists
in the weak coupling
limit only if $\delta\mu/\Delta_{\rm 2SC}$ is held fixed, within
the window (\ref{LOFFwindow}), while the standard weak-coupling limit
$\Delta_{\rm 2SC}/\mu\rightarrow 0$ is taken.  Looking  ahead to our context,
and recalling that in three-flavor quark matter $\delta\mu=M_s^2/(8\mu)$,
we see that at high densities one finds
the CFL phase (which is the three-flavor quark matter BCS phase) and
in some window of lower densities one finds a crystalline phase.
In the vicinity of the second order transition, where $\Delta\rightarrow 0$
and in particular where $\Delta/\delta\mu\rightarrow 0$ and, consequently given (\ref{LOFFwindow}),
$\Delta/\Delta_{\rm 2SC}\rightarrow 0$ a Ginzburg-Landau expansion
of the free energy order by order in powers of $\Delta$ is controlled.

The Ginzburg-Landau
analysis can then be applied to more complicated crystal structures in which
Cooper pairs with several different ${\bf q}$'s, all with the same length but pointing
in different directions, arise.\cite{Bowers:2002xr}  This analysis indicates that a face-centered cubic
structure constructed as the sum of eight plane waves with ${\bf q}$'s pointing at
the corners of a cube is favored, but it does not permit a quantitative evaluation of
$\Delta(\delta\mu)$. The Ginzburg-Landau expansion of the free energy has
terms that are quartic and sextic in $\Delta$ whose coefficients are both
large in magnitude and negative. To this order, $\Omega$ is not bounded from below.
This means that the Ginzburg-Landau analysis
predicts a strong first order phase transition
between the crystalline and unpaired phase, at some $\delta\mu$ significantly larger
than $0.754\, \Delta_{\rm 2SC}$, meaning that the crystalline
phase occurs over a range of $\delta\mu$ that is
much wider than (\ref{LOFFwindow}), but it precludes the 
quantitative evaluation of the $\delta\mu$ at 
which the transition occurs, of $\Delta$, or of $\Omega$.  

In three-flavor quark matter, all the crystalline phases analyzed in
Ref.~\refcite{Rajagopal:2006ig} have Ginzburg-Landau free energies
with positive sextic coefficient, meaning that they can be used to
evaluate $\Delta$, $\Omega$ and the location of the transition from
unpaired quark matter to the crystalline phase with a postulated
crystal structure.  For the most favored crystal structures, we
shall see that the window in parameter space in which they occur
is given by (\ref{WideLOFFWindow}), which is in no sense narrow.

\subsubsection{Crystalline color superconductivity in
  neutral three-flavor quark matter}
\label{sec:LOFF3flav}

We now turn to three-flavor crystalline color superconductivity.  We
shall make weak coupling (namely $\Delta_0,\delta\mu \ll \mu$) and
Ginzburg-Landau (namely $\Delta\ll\Delta_0,\delta\mu$) approximations
throughout.  The analysis of neutrality in three-flavor quark matter
in a crystalline color superconducting phase is very simple in the
Ginzburg-Landau limit in which $\Delta\ll \delta\mu$: because the
construction of this phase does {\it not} involve rearranging any
Fermi momenta prior to pairing, and because the assumption $\Delta\ll
\delta\mu$ implies that the pairing does not significantly change any
number densities, neutrality is achieved with the same chemical
potentials $\mu_e=M_s^2/(4\mu)$ and $\mu_3=\mu_8=0$ as in unpaired
quark matter, and with Fermi momenta given in Eqs.~(\ref{pF1}) and
(\ref{pF3}) as in unpaired quark matter.

We consider a  condensate of the form
\begin{equation}
\langle\psi_{i\alpha}(x) C\gamma_5 \psi_{j\beta}(x)\rangle \propto 
 \sum_{I=1}^3\ \ 
\sum_{\q{I}{a}\in\setq{I}{}}
 \Delta_I  e^{2i\q{I}{a}\cdot\rr}\coleps\flaeps 
 \label{condensate}\;,
\end{equation}
where $\alpha,\beta$ are color indices, $i,j$ are flavor indices, and
where ${\bf q}_1^a$,  ${\bf q}_2^a$ and ${\bf q}_3^a$  and $\Delta_1$, $\Delta_2$  and
$\Delta_3$ are the wave vectors and gap parameters describing pairing between
the $(d,s)$, $(u,s)$ and $(u,d)$ quarks respectively, whose Fermi momenta are
split by $2\delta\mu_1$, $2\delta\mu_2$ and $2\delta\mu_3$ respectively.
{}From (\ref{pF3}), we see that $\delta\mu_2=\delta\mu_3=\delta\mu_1/2=M_s^2/(8\mu)$.
For each $I$, $\setq{I}{}$ is a set of momentum vectors that define the periodic spatial
modulation of the crystalline condensate describing pairing between the quarks
whose flavor is not $I$, and whose color is not $I$.  Our goal in this paper is to
compare condensates with different choices of $\setq{I}{}$'s, that is with different
crystal structures.   To shorten expressions, we will henceforth write
$\sum_{\q{I}{a}}\equiv
\sum_{\q{I}{a}\in\setq{I}{}}$.
The condensate (\ref{condensate}) has the color-flavor structure of the 
CFL condensate (obtained by setting all ${\bf q}$'s to zero) and is the
natural generalization to nontrivial crystal structures of the condensate
previously analyzed in Refs.~\refcite{Casalbuoni:2005zp,Mannarelli:2006fy}, in
which each $\setq{I}{}$ contained only a single vector. 

In all our results, although not in the derivation of the Ginzburg-Landau
approximation itself in Ref.~\refcite{Rajagopal:2006ig},
we shall make the further simplifying assumption
that $\Delta_1=0$.  Given that $\delta\mu_1$ is twice $\delta\mu_2$ or $\delta\mu_3$,
it seems reasonable that $\Delta_1\ll \Delta_2,\Delta_3$.   We leave a quantitative
investigation of condensates with $\Delta_1\neq 0$ to future work.

\subsubsection{NJL Model and Mean-Field Approximation}

As discussed in Section \ref{sec:LOFFintro}, 
we shall work in a NJL model in which the quarks interact
via a point-like four-quark interaction, with the quantum numbers of single-gluon exchange,
analyzed in mean field theory.  By this we mean that the interaction term 
is
\begin{equation}
{\cal L}_{\rm interaction} = 
-\frac{3}{8}\lambda(\bar{\psi}\Gamma^{A\nu}\psi)(\bar{\psi}\Gamma_{A\nu}\psi)
\label{interactionlagrangian}\ ,
\end{equation}
The full expression for $\Gamma^{A\nu}$ is 
$(\Gamma^{A\nu})_{\alpha i,\beta j} = \gamma^\nu (T^A)_{\alpha \beta}\delta_{i j}$, 
where the $T^A$ are the 
color Gell-Mann matrices. The NJL coupling constant $\lambda$ has dimension -2,
meaning that an ultraviolet cutoff $\Lambda$ 
must be introduced as a second parameter in order
to fully specify the interaction.    Defining $\Lambda$ as the restriction that momentum
integrals be restricted to a shell around the Fermi surface, 
$\mu-\Lambda < |{\bf p}| < \mu + \Lambda$,  the CFL gap parameter
can then be evaluated:\cite{Reviews,Bowers:2002xr}
\beq
\Delta_0=2^{\frac{2}{3}}\Lambda \exp\left[-\frac{\pi^2}{2\mu^2\lambda}\right]\ .
\end{equation}
In the limit in which
which $\Delta\ll \Delta_0,\delta\mu\ll\mu$, all our results can be expressed in terms of $\Delta_0$;
neither $\lambda$ nor $\Lambda$ appear.\cite{Rajagopal:2006ig}  This reflects the fact that in this
limit the physics of interest is dominated by quarks near the Fermi surfaces,  not near 
$\Lambda$, and so once $\Delta_0$ is used
as the parameter describing the strength of the attraction between quarks, $\Lambda$ 
is no longer visible; the cutoff $\Lambda$ only appears in the relation between
$\Delta_0$ and $\lambda$, not in any comparison among different possible
paired phases.
In our numerical evaluations,
we shall take $\mu=500$~MeV, $\Lambda=100$~MeV,  and adjust $\lambda$ to be
such that $\Delta_0$ is $25$~MeV. 

In the mean-field approximation, the interaction Lagrangian (\ref{interactionlagrangian})
takes the form
\begin{equation}
{\cal L}_{\rm interaction}=
\ha\bar{\psi}\Delta(x)\bar{\psi}^T + \ha\psi^T\bar{\Delta}(x)\psi,
\label{meanfieldapprox}
\end{equation}
where $\Delta(x)$ is related to the diquark condensate by the
relations
\begin{equation}
\begin{split}
\Delta(x) &= \frac{3}{4}\lambda\Gamma^{A\nu}\langle\psi\psi^T\rangle(\Gamma_{A\nu})^T \ .
\end{split}
\end{equation}
The ansatz (\ref{condensate}) can now be made precise: we take
\begin{equation}
\Delta(x)=\Delta_{CF}(x)\otimes C\gamma^5\label{spin structure}\;,
\end{equation}
with 
\begin{equation}
\Delta_{CF}(x)_{\cf} = \sum_{I=1}^3\sum_{\q{I}{a}}
 \Delta_I e^{2i\q{I}{a}\cdot \rr}\coleps\flaeps\label{precisecondensate}\ .
\end{equation}
Upon making the mean field approximation, the full Lagrangian
is quadratic and can be written simply upon introducing two component
Nambu-Gorkov spinors. Upon so doing, gap equations can easily be 
derived.\cite{Rajagopal:2006ig}
Without further approximation, however, they are not tractable because
without further approximation it is not consistent to choose finite sets $\{{\bf q}_I\}$. 
When several plane waves are present in the condensate, they induce an infinite
tower of higher momentum condensates.\cite{Bowers:2002xr}   The reason why
the Ginzburg-Landau approximation, to which we now turn, is such a simplification
is that it eliminates these higher harmonics. 

\subsection{Ginzburg-Landau Approximation}

The form of the Ginzburg-Landau expansion of the free energy can be derived using only
general arguments.  

We shall only consider crystal structures in which
all the vectors $\q{I}{a}$ in the crystal structure $\{ {\bf q}_I\}$ are ``equivalent''.
By this we mean that a rigid rotation of the crystal structure can be found which
maps any $\q{I}{a}$ to any other $\q{I}{b}$ leaving the set $\{ {\bf q}_I \}$ invariant.
(If we did not make this simplifying assumption, we would need to introduce
different gap parameters $\Delta_I^a$ for each vector in $\setq{I}{}$.)
As explained in Section \ref{sec:LOFF3flav}, 
the chemical potentials that maintain neutrality in
three-flavor crystalline color superconducting quark matter are the same as those
in neutral unpaired three-flavor quark matter.  Therefore, 
\beq
\Omega_{\rm crystalline}=\Omega_{\rm unpaired} + \Omega(\Delta_1,\Delta_2,\Delta_3)\ ,
\label{CondensationEnergy}
\eeq
with $\Omega_{\rm unpaired}$ given in (\ref{OmegaUnpaired}) with (\ref{mueneutral}), 
and 
with $\Omega(0,0,0)=0$. Our task is to evaluate the condensation energy 
$\Omega(\Delta_1,\Delta_2,\Delta_3)$.   Since our Lagrangian is baryon number
conserving and 
contains no
weak interactions, it is invariant under a global $U(1)$ symmetry for each flavor.
This means that $\Omega$ must be invariant under $\Delta_I \rightarrow e^{i\phi_I} \Delta_I$ 
for each $I$, meaning that each of the three $\Delta_I$'s can only appear in the combination
$\Delta_I^*\Delta_I$.  (Of course, the ground state can and does break these
$U(1)$ symmetries spontaneously; what we  need in the argument we are making here is only
that they are not explicitly broken in the Lagrangian.)  We conclude that if 
we expand $\Omega(\Delta_1,\Delta_2,\Delta_3)$ in powers of the $\Delta_I$'s up
to sextic order, it must take the form
\begin{equation}
\begin{split}
\Omega&(\{\Delta_I\})= \frac{2\mu^2}{\pi^2}\Biggl[\sum_I P_I
\alpha_I \, \Delta_I^*\Delta_I +\ha\Biggl(\sum_I \beta_I(\Delta_I^*\Delta_I)^2
+\sum_{I>J} \beta_{IJ}\, \Delta_I^*\Delta_I\Delta_J^*\Delta_J\Biggr)\\
&+\frac{1}{3}\Biggl(\sum_I \gamma_I(\Delta_I^*\Delta_I)^3
+\sum_{I\neq J}\gamma_{IJJ}\,
\Delta_I^*\Delta_I\Delta_J^*\Delta_J\Delta_J^*\Delta_J
+\gamma_{123}\,\Delta_1^*\Delta_1\Delta_2^*\Delta_2\Delta_3^*\Delta_3\Biggr)\Biggr]
\label{GLexpansion}\;,
\end{split}
\end{equation}
where we have made various notational choices for convenience.  The overall prefactor
of $2\mu^2/\pi^2$ is the density of states at the Fermi surface of unpaired quark 
matter with $M_s=0$; it is convenient to define all the coefficients in
the Ginzburg-Landau expansion of the free energy relative to this.  We have defined
$P_I={\rm dim}\{{\bf q}_I\}$, the number of plane  waves in the crystal structure
for the condensate describing pairing between quarks whose flavor and color are
not $I$.  Writing the prefactor $P_I$ multiplying the quadratic term 
and writing the factors of $\ha$ and $\frac{1}{3}$ multiplying the 
quartic and sextic terms  ensures that the $\alpha_I$, $\beta_I$ and $\gamma_I$ coefficients
are defined the same way as in Ref.~\refcite{Bowers:2002xr}.
The form of the
Ginzburg-Landau expansion (\ref{GLexpansion}) is model independent, whereas
the expressions for the coefficients $\alpha_I$, $\beta_I$, $\beta_{IJ}$, $\gamma_I$,
$\gamma_{IJJ}$ and $\gamma_{123}$ for a given ansatz
for the crystal structure  are model-dependent. 
In Sections IV
and V of Ref.~\refcite{Rajagopal:2006ig} these coefficients are
evaluated by deriving the Ginzburg-Landau approximation to this model.

We see in  Eq.~(\ref{GLexpansion}) that there are some coefficients --- namely
$\alpha_I$, $\beta_I$ and $\gamma_I$ --- which multiply polynomials involving
only a single $\Delta_I$.   Suppose that we keep a single $\Delta_I$ nonzero,
setting the other two to zero.  This reduces the problem to one with two-flavor
pairing only, and the Ginzburg-Landau coefficients for this problem have been
calculated for many different crystal structures in Ref.~\refcite{Bowers:2002xr}.
We can then immediately use these coefficients, called $\alpha$, $\beta$
and $\gamma$ in Ref.~\refcite{Bowers:2002xr}, to determine our $\alpha_I$, 
$\beta_I$ and $\gamma_I$.  Using $\alpha_I$ as an example, we conclude
that
\begin{equation}
\begin{split}
\alpha_I&=
\alpha(q_I,\delta\mu_I)= -1
 +\frac{\delta\mu_I}{2 q_I}
 \log\left(\frac{q_I+\delta\mu_I}{q_I-\delta\mu_I}\right)
 -\ha\log\left(\frac{\Delta_{\rm 2SC}^2}{4(q_I^2-\delta\mu_I^2)}\right)\ ,
\label{AlphaEqn}
\end{split}
\end{equation}
where $\delta\mu_I$ is the splitting between the Fermi surfaces of the quarks
with the two flavors other than $I$ and $q_I\equiv |\q{I}{a}|$ is the length of
the ${\bf q}$-vectors in the set $\{{\bf q}_I\}$.  (We shall see momentarily why
all have the same length.)
In (\ref{AlphaEqn}), $\Delta_{\rm 2SC}$ is the gap parameter
in the BCS state obtained with $\delta\mu_I=0$ and $\Delta_I$ nonzero with the
other two gap parameters set to zero.    Assuming that $\Delta_0\ll \mu$, this gap
parameter for 2SC (2-flavor, 2-color) BCS pairing is given by\cite{Schafer:1999fe,Reviews}
\beq
\Delta_{\rm 2SC}= 2^{\frac{1}{3}}\Delta_0\ .
\eeq
In the Ginzburg-Landau approximation, in which the $\Delta_I$ are assumed
to be small, we must  first minimize the quadratic contribution to the
free energy, before proceeding to investigate the consequences of
the quartic and sextic contributions.  
Notice that $\alpha_I$ 
depends on the cutoff $\Lambda$ and the
NJL coupling constant $\lambda$ only through $\Delta_{\rm 2SC}$, and depends
only on the ratios $q_I/\delta\mu_I$ and $\delta\mu_I/\Delta_{\rm 2SC}$.
$\alpha_I$ is negative for $\delta\mu_I/\Delta_{\rm 2SC}<0.754$,
and for a given value of this ratio for which $\alpha_I<0$, $\alpha_I$ 
is most negative
and (\ref{AlphaEqn}) is minimized 
for\cite{LOFF,Alford:2000ze,Bowers:2002xr}
\beq
q_I = \eta\, \delta\mu_I ~{\rm with}~ \eta=1.1997 \ ,
\label{EtaEqn}
\eeq
where $\eta$ is defined as the solution to
\beq
\frac{1}{2\eta}\log\left(\frac{\eta+1}{\eta-1}\right)=1\ .
\label{EtaDefn}
\eeq
We therefore set $q_I=\eta\,\delta\mu_I$ henceforth and upon so doing find that
(\ref{AlphaEqn}) becomes
\beq
\alpha(\delta\mu_I) 
=-\frac{1}{2}
\log\left(\frac{\Delta_{\rm 2SC}^2}{4 \delta\mu_I^2 (\eta^2-1)}\right)\ .
\label{AlphaEqn2}
\eeq
Minimizing $\alpha_I$ has fixed the length of all the ${\bf
q}$-vectors in the set $\{{\bf q}_I\}$, thus eliminating the
possibility of higher harmonics.

It is helpful to imagine the (three) sets $\setq{I}{}$ as representing
the vertices of (three) polyhedra in momentum space.  By minimizing
$\alpha_I$, we have learned that each polyhedron $\setq{I}{}$ can be
inscribed in a sphere of radius $\eta \delta\mu_I$.  From the
quadratic contribution to the free energy, we do not learn anything
about what shape polyhedra are preferable.  In fact, the quadratic
analysis in isolation would indicate that if $\alpha_I<0$ (which
happens for $\delta\mu_I<0.754\,\Delta_{\rm 2SC}$) then modes with
arbitarily many different $\hat{\bf q}_I$'s should condense.  It is
the quartic and sextic coefficients that describe the interaction
among the modes, and hence control what shape polyhedra are in fact
preferable.

The quartic and sextic coefficients $\beta_I$ and $\gamma_I$ can also
be taken directly from the two-flavor results of
Ref.~\refcite{Bowers:2002xr}.  Note that these coefficients are
independent of $\Delta_{\rm 2SC}$ and the NJL coupling constant and
cutoff, as long as the weak-coupling approximation $\Delta_{\rm
2SC}\ll \mu$ is valid.\cite{Bowers:2002xr,Rajagopal:2006ig}
Given this, and given (\ref{EtaEqn}), the only dimensionful quantity
on which they can depend is $\delta\mu_I$.  They are therefore given
by $\beta_I=\bar\beta/\delta\mu_I^2$ and
$\gamma_I=\bar\gamma/\delta\mu_I^4$ where $\bar\beta_I$ and
$\bar\gamma_I$ are dimensionless quantities depending only on the
directions of the vectors in the set $\setq{I}{}$. $\bar\beta$ and
$\bar\gamma$ have been evaluated for many crystal structures in
Ref.~\refcite{Bowers:2002xr}, resulting in two qualitative
conclusions.  Recall that, as reviewed in Section \ref{sec:LOFF2flav},
the presence of a condensate with some $\hat{\bf q}_I^a$ corresponds
to pairing on a ring on each Fermi surface with opening angle
$67.1^\circ$.  The first qualitative conclusion is that any crystal
structure in which there are two $\hat{\bf q}_I^a$'s whose pairing
rings intersect has {\it very} large, positive, values of both
$\bar\beta$ and $\bar\gamma$, meaning that it is strongly disfavored.
The second conclusion is that regular structures, those in which there
are many ways of adding four or six $\hat{\bf q}_I^a$'s to form closed
figures in momentum space, are favored.  Consequently, the favored
crystal structure in the two-flavor case has 8 $\hat{\bf q}_I^a$'s
pointing towards the corners of a cube.\cite{Bowers:2002xr} Choosing
the polyhedron in momentum space to be a cube yields a face-centered
cubic modulation of the condensate in position space.

Because the $\beta_I$ and $\gamma_I$ coefficients in our problem can
be taken over directly from the two-flavor analysis, we can expect
that it will be unfavorable for any of the three sets $\setq{I}{}$ to
have more than eight vectors, or to have any vectors closer together
than $67.1^\circ$.  It can be proved that $\beta_{IJ}$ is always
positive, and in all cases investigated to date $\gamma_{IIJ}$ also
turns out to be positive.\cite{Rajagopal:2006ig} This means that we
know of no exceptions to the rule that if a particular $\setq{I}{}$ is
unfavorable as a two-flavor crystal structure, then any three-flavor
condensate in which this set of ${\bf q}$-vectors describes either the
$\De_1$, $\De_2$ or $\De_3$ crystal structure is also disfavored.

The coefficients $\beta_{IJ}$ and $\gamma_{IIJ}$ cannot be read off
from a two-flavor analysis because they multiply terms involving more
than one $\Delta_I$ and hence describe the interaction between the
three different $\Delta_I$'s.  Before evaluating the expressions for
the coefficients,\cite{Rajagopal:2006ig} we make the further
simpifying assumption that $\Delta_1=0$, because the separation
$\delta\mu_1$ between the $d$ and $s$ Fermi surfaces is twice as large
as that between either and the intervening $u$ Fermi surface.  This
simplifies (\ref{GLexpansion}) considerably, eliminating the
$\gamma_{123}$ term and all the $\beta_{IJ}$ and $\gamma_{IIJ}$ terms
except $\beta_{32}$, $\gamma_{223}$ and $\gamma_{332}$.
We further simplify the problem by focussing on crystal structures for
which $\{\hat{\bf q}_2\}$ and $\{\hat{\bf q}_3\}$ are ``exchange
symmetric'', meaning that there is a sequence of rigid rotations and
reflections which when applied to all the vectors in $\setq{2}{}$ and
$\setq{3}{}$ together has the effect of exchanging $\{\hat{\bf q}_2\}$
and $\{\hat{\bf q}_3\}$.  If we choose an exchange symmetric crystal
structure, upon making the approximation that
$\delta\mu_2=\delta\mu_3$ and restricting our attention to solutions
with $\Delta_2=\Delta_3$ we have the further simplification that
$\gamma_{322}=\gamma_{233}$.

Upon making these simplifying assumptions, the only dimensionful 
quantity on which $\beta_{32}$ and $\gamma_{322}$ can depend
is $\delta\mu$, meaning that
$\beta_{32}=\bar\beta_{32}/\delta\mu^2$ and
$\gamma_{322}=\bar\gamma_{322}/\delta\mu^4$.\cite{Rajagopal:2006ig}
Here, $\bar\beta_{32}$ and $\bar\gamma_{322}$ 
are dimensionless numbers, depending only on the shape and relative
orientation of the polyhedra $\{\hat{\bf q}_2\}$ 
and $\{\hat{\bf q}_3\}$, which must be evaluated for each
crystal structure.  Doing so requires evaluating one loop Feynman
diagrams with 4 or 6 insertions of $\Delta_I$'s. Each insertion of
$\Delta_I$ ($\Delta_I^*$) adds (subtracts) momentum $2\q{I}{a}$ for
some $a$, meaning that the calculation consists of a bookkeeping task
(determining which combinations of 4 or 6 $\q{I}{a}$'s are allowed)
that grows rapidly in complexity with the compexity of the crystal
structure and a loop integration that is nontrivial because the
momentum in the propagator changes after each insertion. See
Ref.~\refcite{Rajagopal:2006ig}, where this calculation is carried out
explicitly for 11 crystal structures in a mean-field NJL model upon
making the weak coupling ($\Delta_0,\delta\mu\ll\mu$) approximation. 
Within this approximation, neither the
NJL cutoff nor the NJL coupling constant appear in any quartic or
higher Ginzburg-Landau coefficient, and they influence $\alpha$ only
through $\Delta_{\rm 2SC}\propto \Delta_0$.  
Hence, the details of the model do not matter as
long as one thinks of $\Delta_0$ as a parameter, kept $\ll \mu$.

\subsection{General results}
\label{sec:LOFFgeneral}

Upon assuming that $\Delta_1=0$, assuming that the crystal structure
is exchange symmetric,
and making the approximation that
$\delta\mu_2=\delta\mu_3\equiv \delta\mu = M_s^2/(8\mu)$,
the free energy (\ref{GLexpansion})
reduces to
\begin{equation}
\begin{split}
\Omega(\Delta_2,\Delta_3)=\frac{2{\mu}^2}{\pi^2}\Biggl[&
P\alpha(\delta\mu)\bigl(\Delta_2^2+\Delta_3^2\bigr)
+\ha\frac{1}{\delta\mu^2}\bigl(\bar{\beta}(\Delta_2^4+\Delta_3^4)
 +\bar{\beta}_{32}\Delta_2^2\Delta_3^2\bigr)\\
 &\!\!\!\!\!+\frac{1}{3}\frac{1}{\delta\mu^4}\bigl(\bar{\gamma}(\Delta_2^6+\Delta_3^6)
 +\bar{\gamma}_{322}(\Delta_2^2\Delta_3^4+\Delta_2^4\Delta_3^2)\bigr)\Biggr]
 \label{congruent equal free energy}\;,
\end{split}
\end{equation}
where $\bar\beta$, $\bar\gamma$, $\bar\beta_{32}$ and $\bar\gamma_{322}$ 
are the dimensionless constants that can be calculated for each crystal 
structure,\cite{Rajagopal:2006ig} and
where  the $\delta\mu$-dependence of $\alpha$ is given by Eq.~(\ref{AlphaEqn2}).
It is easy to show that  
extrema of $\Omega(\Delta_2,\Delta_3)$ in $(\Delta_2,\Delta_3)$-space
must either have $\Delta_2=\Delta_3=\Delta$,
or have one of $\Delta_2$ and $\Delta_3$ vanishing.\cite{Rajagopal:2006ig}  
The latter class of extrema are
two-flavor crystalline phases.  We are interested in the solutions with 
$\Delta_2=\Delta_3=\Delta$.   
(Furthermore,
the three-flavor crystalline phases with 
$\Delta_2=\Delta_3=\Delta$ are electrically neutral whereas the two-flavor solutions
in which only one of the $\Delta$'s is nonzero are not.\cite{Rajagopal:2006ig})  
Setting $\Delta_2=\Delta_3=\De$, the free energy becomes
\begin{equation}
\Omega(\Delta) =  \frac{2{\mu}^2}{\pi^2}
 \left[2P\alpha(\delta\mu)\Delta^2
+\frac{\Delta^4}{2\delta\mu^2}\bar{\beta}_{\rm eff}
+\frac{\Delta^6}{3\delta\mu^4}\bar{\gamma}_{\rm eff}\right]
 \label{effective free energy}\;, 
\end{equation}
where we have defined
\beq
\begin{split}
\bar{\beta}_{\rm eff}&=2\bar{\beta} + \bar{\beta}_{32}\\
\bar{\gamma}_{\rm eff}&=2\bar{\gamma}+2\bar{\gamma}_{322}\ .
\end{split}
\label{BetaGammaEff}
\eeq
We have arrived at a familiar-looking sextic order Ginzburg-Landau free energy
function, whose coefficients can be evaluated for specific crystal structures.\cite{Rajagopal:2006ig}

If $\bar{\beta}_{\rm eff}$ and $\bar{\gamma}_{\rm eff}$ are both positive,
the free energy (\ref{effective free energy}) describes a second order
phase transition between the crystalline color superconducting phase
and the normal phase at the $\delta\mu$ at which $\alpha(\delta\mu)$
changes sign.  From (\ref{AlphaEqn2}), this critical point occurs
where $\delta\mu=0.754\, \Delta_{\rm 2SC}$.  In plotting our results,
we will take the CFL gap to be $\Delta_0=25$~MeV,
making $\Delta_{\rm 2SC}=2^{1/3}\Delta_0=31.5$~MeV.
Recalling that $\delta\mu=M_s^2/(8\mu)$,
this puts the second order phase transition at 
\beq
\frac{M_s^2}{\mu}\Biggr|_{\rm \alpha=0}=6.03\, \Delta_{\rm 2SC} = 7.60\,\Delta_0 = 190.0~{\rm MeV}\ .
\label{AlphaZero}
\eeq
(The authors of Refs.~\refcite{Casalbuoni:2005zp,Mannarelli:2006fy} 
neglected to notice that it is $\Delta_{\rm 2SC}$, rather
than the CFL gap $\Delta_0$, that occurs in Eqs.~(\ref{AlphaEqn}) 
and (\ref{AlphaEqn2}) and therefore controls 
the $\delta\mu$ at 
which $\alpha=0$. 
In analyzing the crystalline phase in isolation, this
is immaterial since either $\Delta_0$ or $\Delta_{\rm 2SC}$ could be taken as
the parameter defining the strength of the interaction between quarks.  However,
we shall compare the free energies of the CFL, gCFL and crystalline
phases, and in making this comparison it is important to take into account that
$\Delta_{\rm 2SC}=2^{1/3}\Delta_0$.)  For values of $M_s^2/\mu$ that
are smaller than (\ref{AlphaZero}) (that is, lower densities), $\alpha<0$ and
the free energy is minimized by a nonzero $\Delta=\Delta_{\rm min}$ 
and thus describes a crystalline color superconducting phase.

If $\bar\beta_{\rm eff}<0$ and $\bar\gamma_{\rm eff}>0$, then the free energy
(\ref{effective free energy}) describes a first order
phase transition between unpaired and crystalline quark matter occurring
at 
$\alpha = \alpha_* =  3\,\bar\beta^2_{\rm eff}/(32\, P \bar\gamma_{\rm eff})$.
At this positive value of $\alpha$, the function $\Omega(\Delta)$ has a minimum at $\Delta=0$
with $\Omega=0$, 
initially rises quadratically with increasing $\Delta$,
and is then turned back downward by
the negative quartic term before being turned back upwards again
by the positive sextic term, yielding a second
minimum at
\beq
\Delta=\delta\mu \sqrt{ \frac{3\, |\bar\beta_{\rm eff }|}{4\,\bar\gamma_{\rm eff}}}\ ,
\label{DeltaAlphaStar}
\eeq
also with $\Omega=0$, which describes a crystalline color superconducting
phase.
For $\alpha<\alpha_*$, the crystalline phase is favored over unpaired quark matter.
Eq.~(\ref{AlphaEqn2}) determines the value of $\delta\mu$, and
hence $M_s^2/\mu$, at which 
$\alpha=\alpha_*$ and the first order phase transition occurs.  If $\alpha_*\ll 1$,
the transition occurs at a value of $M_s^2/\mu$ that is greater than (\ref{AlphaZero}) by
a factor $(1+\alpha_*)$.  See Fig.~\ref{OmegaVsDeltaFig} in Section \ref{sec:LOFFfree}
for an explicit example of plots of $\Omega$ versus $\Delta$
for various values of $\alpha$ for one of the crystal structures that we describe.

A necessary condition for the Ginzburg-Landau approximation to be quantitatively
reliable is that the sextic term in the free energy is small
in magnitude compared to the quartic, meaning
that $\De^2\ll \dmu^2 |\bar\beta_{\rm eff}/\bar\gamma_{\rm eff}|$.  If the transition
between the unpaired and crystalline phases
is second order, then this condition is satisfied close enough to the transition
where $\De\rightarrow 0$.  However, if $\bar\beta_{\rm eff}<0$ and $\bar\gamma_{\rm eff}>0$,
making the transition first order, we see from (\ref{DeltaAlphaStar}) that at the
first order transition itself $\Delta$ is large enough to make the quantitative
application of the Ginzburg-Landau approximation marginal.  This is a familiar 
result, coming about whenever a Ginzburg-Landau approximation predicts
a first order phase transition
because at the first order phase transition the quartic
and sextic terms are balanced against each other.   
We shall find that our Ginzburg-Landau analysis predicts a first
order phase transition; knowing that it is therefore at the edge
of its quantitative reliability, we shall focus on qualitative
conclusions.

\subsection{Two plane wave structure}
\label{sec:LOFF2plane}

We begin with the simplest three-flavor ``crystal'' structure in which
$\setq{2}{}$ and $\setq{3}{}$ each contain only a single vector,
yielding a condensate in which the $\langle us\rangle$ and $\langle
ud\rangle$ condensates are each plane waves.  This simple condensate
yields a qualitative lesson which proves extremely helpful in
winnowing the space of multiple plane wave crystal
structures.\cite{Rajagopal:2006ig}

For this simple ``crystal'' structure, 
all the coeffficients in the Ginzburg-Landau
free energy can be evaluated analytically.\cite{Rajagopal:2006ig} The 
terms that occur in the three-flavor case but not in the two-flavor case,
namely $\bar\beta_{32}$ and $\bar\gamma_{322}$, describe the
interaction between the two condensates and hence depend on the angle
$\phi$ between $\q{2}{}$ and $\q{3}{}$. 
For any angle $\phi$, both $\bar{\beta}_{32}$ and $\bar{\gamma}_{322}$ are
positive,
both increase monotonically with $\phi$, and both
diverge as $\phi\rightarrow\pi$.\cite{Rajagopal:2006ig}  This tells us that within this two plane
wave ansatz, the most favorable orientation is $\phi=0$, namely
$\q{2}{}\parallel\q{3}{}$.  Making this choice yields the smallest possible
$\bar\beta_{\rm eff}$ and $\bar\gamma_{\rm eff}$ within this ansatz, and
hence the largest possible $\Delta$ and condensation energy, again within
this ansatz.
The divergence at $\phi\rightarrow\pi$
tells us that choosing $\q{2}{}$ and
$\q{3}{}$ precisely antiparallel exacts an infinite free energy price
in the combined Ginzburg-Landau and weak-coupling
limit in which $\Delta\ll \delta\mu,\Delta_0\ll \mu$,
meaning that in this limit if we chose $\phi=\pi$ we find $\Delta=0$.
Away from the Ginzburg-Landau limit, when the pairing rings
on the Fermi surfaces widen into bands, choosing $\phi=\pi$ exacts a finite
price meaning that $\Delta$  is nonzero but smaller than that for any other
choice of $\phi$.  

The high cost of choosing $\q{2}{}$ and
$\q{3}{}$ precisely antiparallel can be understood 
qualitatively as arising from the fact that in this case the
ring of states on the $u$-quark Fermi surface that ``want to'' pair
with $d$-quarks coincides precisely with the ring that ``wants to''
pair with $s$-quarks.\cite{Mannarelli:2006fy}
The simple two plane wave ansatz has been
analyzed in the same NJL model that we employ upon making
the weak-coupling approximation but without
making a Ginzburg-Landau approximation.\cite{Mannarelli:2006fy}
All the qualitative lessons that we have learned from the Ginzburg-Landau approximation
remain valid, including the favorability of the choice $\phi=0$,
but we learn further that in the two plane wave case
the Ginzburg-Landau approximation always underestimates $\Delta$.\cite{Mannarelli:2006fy}

The analysis of the simple two plane wave ``crystal'' structure, together with
the observation that in more complicated crystal structures with more than
one vector in $\setq{2}{}$ and $\setq{3}{}$ the Ginzburg-Landau coefficient
$\beta_{32}$ ($\gamma_{322}$) is given in whole (in part) by a sum of many 
two plane wave contributions,
yields an important lesson for how to construct favorable crystal structures for three-flavor
crystalline color superconductivity:\cite{Rajagopal:2006ig}
$\setq{2}{}$ and $\setq{3}{}$ should be rotated with respect to each other
in a way that best keeps vectors in one set away from the antipodes
of vectors in the other set.  
Crystal structures in which any of the
vectors in $\setq{2}{}$ are close  to antiparallel to any of the vectors
in $\setq{3}{}$ are strongly disfavored because if a vector 
in $\setq{2}{}$ is antiparallel (or close to antiparallel) to one in $\setq{3}{}$,
this yields infinite (or merely large) positive contributions to $\bar\beta_{32}$ and
to $\bar\gamma_{322}$ and hence to $\bar\beta_{\rm eff}$
and $\bar\gamma_{\rm eff}$, meaning that there is
an infinite (or large) free energy penalty for $\Delta\neq 0$.   

Summarizing, we have arrived at two rules for constructing favorable
crystal structures for three-flavor crystalline color
superconductivity.  First, the sets $\setq{2}{}$ and $\setq{3}{}$
should each be chosen to yield crystal structures which, seen as
separate two-flavor crystalline phases, are as favorable as possible.
In Section \ref{sec:LOFF2flav} we have reviewed how this should be
done and the conclusion that the most favored $\setq{2}{}$ or
$\setq{3}{}$ in isolation consists of eight vectors pointing at the
corners of a cube.\cite{Bowers:2002xr} Second, the new addition in the
three-flavor case is the qualitative principle that $\setq{2}{}$ and
$\setq{3}{}$ should be rotated with respect to each other in such a
way as to best keep vectors in one set away from the antipodes of
vectors in the other set.

\subsection{Multiple plane waves}
\label{sec:LOFFmultiple}

Rajagopal and Sharma have analyzed 11 different crystal
structures,\cite{Rajagopal:2006ig} in each case calculating
$\bar\beta$ and $\bar\gamma$ and $\bar\beta_{32}$ and
$\bar\gamma_{322}$, and hence $\bar\beta_{\rm eff}$ and
$\bar\gamma_{\rm eff}$ that specify the free energy as in
(\ref{BetaGammaEff}).  The 11 structures allow one to make many
pairwise comparisons that test the two qualitative principles
described in Section \ref{sec:LOFF2plane}.  There are some instances
of two structures which differ only in the relative orientation of
$\setq{2}{}$ and $\setq{3}{}$ and in these cases the structure in
which vectors in $\setq{2}{}$ get closer to the antipodes of vectors
in $\setq{3}{}$ are always disfavored. And, there are some instances
where the smallest angle between a vector in $\setq{2}{}$ and the
antipodes of a vector in $\setq{3}{}$ are the same for two different
crystal structures, and in these cases the one with the more favorable
two-flavor structure is more favorable.  These considerations,
together with explicit calculations, indicate that two structures,
which we denote ``2Cube45z'' and ``CubeX'', are particularly
favorable.

\begin{figure}[t]
\bc
\includegraphics[width=0.9\hsize]{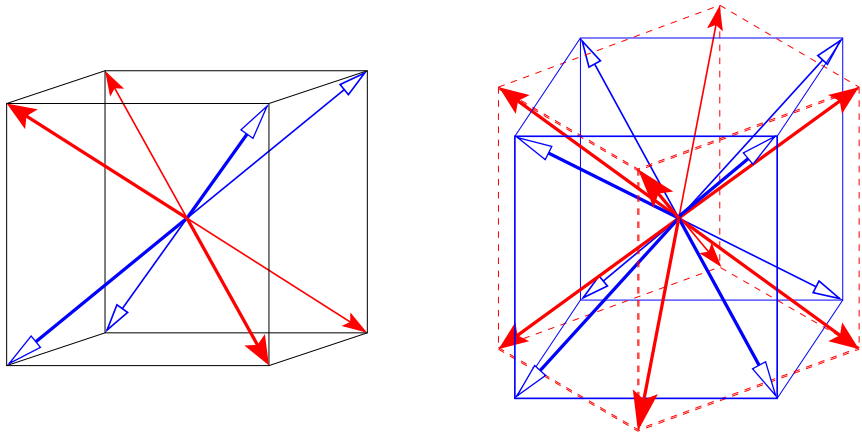}
\ec
\caption{The momenta that constitute the CubeX (left) and 2Cube45z (right)
crystal structures. The set ${\bf q}_3$ of $\<ud\>$ pairing momenta is given
by the solid-tipped (red online) arrows.
The set ${\bf q}_2$ of $\<us\>$ pairing momenta is given by the 
hollow-tipped (blue online) arrows.
}
\label{fig:momenta}
\end{figure}

The 2Cube45z crystal momenta are illustrated in Fig.~\ref{fig:momenta},
right panel;
$\setq{2}{}$ and $\setq{3}{}$ each contain eight
vectors pointing at the corners of a cube. If we orient $\setq{2}{}$ so that its
vectors point in the $(\pm 1,\pm 1,\pm 1)$ directions, then $\setq{3}{}$ is rotated
relative to $\setq{2}{}$ by $45^\circ$ about the $z$-axis.   In this crystal structure,
the $\langle ud \rangle$ and $\langle us \rangle$ condensates are each
face-centered cubic, the structure which in isolation is the most favored 
two-flavor crystal structure.\cite{Bowers:2002xr}  The relative rotation
has been chosen to maximize the separation between any vector in
$\setq{2}{}$ and the nearest antipodes of a vector in $\setq{3}{}$.
The color-flavor
and position space  dependence of the condensate, defined in (\ref{spin structure})
and (\ref{precisecondensate}),
is given by
\begin{equation}
\begin{split}
\Delta_{CF}(x)_{\cf} = & \epsilon_{2\alpha\beta}\epsilon_{2ij} \, 
2\Delta \Biggl[ 
\cos \frac{2\pi}{a} \left( x+y+z\right) + \cos \frac{2\pi}{a} \left(-x+y+z\right)\\
&\qquad\qquad+\cos \frac{2\pi}{a} \left(x-y+z\right) 
+ \cos \frac{2\pi}{a}\left(-x-y+z\right) \Biggr]\\
&+ \epsilon_{3\alpha\beta}\epsilon_{3ij} \, 2\Delta \Biggl[
\cos \frac{2\pi}{a} \left( \sqrt{2} x +z\right) + \cos \frac{2\pi}{a} \left(\sqrt{2} y+z\right)\\
&\qquad\qquad+\cos \frac{2\pi}{a} \left(-\sqrt{2}y+z\right) 
+ \cos \frac{2\pi}{a}\left(-\sqrt{2}x+z\right) \Biggr]\ ,
\label{2Cube45zStructure}
\end{split}
\end{equation}
where $\alpha$ and $\beta$ ($i$ and $j$) are color (flavor) indices and 
where
\beq
a = \frac{\sqrt{3}\pi}{q} = \frac{4.536}{\delta\mu} = \frac{\mu}{1.764 M_s^2}
\label{LatticeSpacing}
\eeq
is the lattice spacing of the face-centered cubic crystal structure.  For
example, with $M_s^2/\mu=100, 150, 200$~MeV the lattice
spacing is  $a=72, 48, 36$~fm.  Eq.~(\ref{2Cube45zStructure}) can equivalently
be written as
$\Delta_{CF}(x)_{\cf} = \epsilon_{2\alpha\beta}\epsilon_{2ij} \Delta_2(\rr)
+ \epsilon_{3\alpha\beta}\epsilon_{3ij} \Delta_3(\rr)$,
with (\ref{2Cube45zStructure}) providing the expressions for $\Delta_2(\rr)$
and $\Delta_3(\rr)$.
A three-dimensional contour
plot that can be seen as depicting 
either $\Delta_2(\rr)$ or $\Delta_3(\rr)$ separately
can be found in Ref.~\refcite{Bowers:2002xr}.
We have not found an informative way of depicting the entire condensate
in a single contour plot.  Note also that in (\ref{2Cube45zStructure})
and below in our description of the CubeX phase, we make an arbitrary
choice for the relative position of $\Delta_3(\rr)$ and $\Delta_2(\rr)$.  We show
in Ref.~\refcite{Rajagopal:2006ig} that one can be translated relative to the other at no cost
in free energy.  Of course, 
rotating one relative to the other changes the
Ginzburg-Landau coefficients $\bar\beta_{32}$ and $\bar\gamma_{322}$ 
and hence the free energy.

\begin{figure}[t]
\bc
\includegraphics[width=0.5\hsize,angle=0]{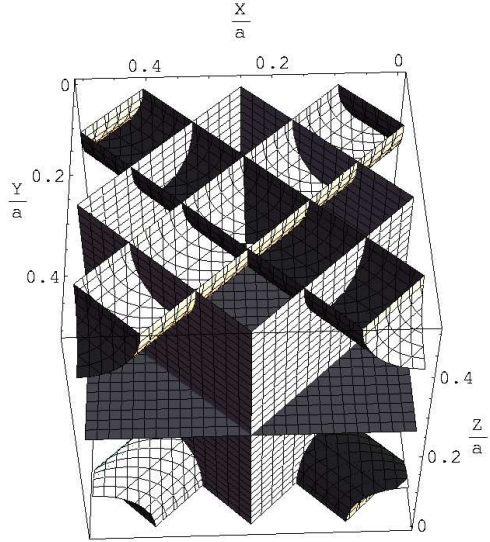}
\ec
\caption{
The CubeX crystal structure of Eq.~(\ref{CubeXStructure}).  The figure
extends from 0 to $a/2$ in the $x$, $y$ and $z$ directions.  Both
$\Delta_2(\rr)$ and $\Delta_3(\rr)$ vanish at the horizontal plane.
$\Delta_2(\rr)$ vanishes on the darker vertical planes, and
$\Delta_3(\rr)$ vanishes on the lighter vertical planes.  On the upper
(lower) dark cylinders and the lower (upper) two small corners of dark
cylinders, $\Delta_2(\rr)= +3.3 \Delta$ ($\Delta_2(\rr)= -3.3
\Delta$).  On the upper (lower) lighter cylinders and the lower
(upper) two small corners of lighter cylinders, $\Delta_3(\rr)= -3.3
\Delta$ ($\Delta_3(\rr)= +3.3 \Delta$).  Note that the largest value
of $|\Delta_I(\rr)|$ is $4\Delta$, occurring along lines at the
centers of the cylinders. The lattice spacing is $a$ when one takes
into account the signs of the condensates; if one looks only at
$|\Delta_I(\rr)|$, the lattice spacing is $a/2$. $a$ is given in
(\ref{LatticeSpacing}).  In (\ref{CubeXStructure}) and hence in this
figure, we have made a particular choice for the relative position of
$\Delta_3(\rr)$ versus $\Delta_2(\rr)$. In fact, one can be translated
relative to the other with no cost in free energy.
}
\label{CubeXContours}
\end{figure}

We now turn to a description of the  CubeX crystal 
structure.  We arrive at this structure by reducing the number
of vectors in $\setq{2}{}$ and $\setq{3}{}$.  This worsens the two-flavor
free energy of each condensate separately, but allows vectors in 
$\setq{2}{}$ to be kept farther away from the antipodes of vectors
in $\setq{3}{}$.  We do not claim to have analyzed all structures
obtainable in this way, but we have found one and only one which
has a condensation energy comparable in magnitude to the 2Cube45z structure.
The CubeX structure is illustrated in Fig.~\ref{fig:momenta},
left panel;
$\setq{2}{}$ and $\setq{3}{}$ each contain four vectors forming a rectangle. 
The eight 
vectors together point toward the corners of a cube. The two rectangles intersect to look like an ``X'' 
if viewed end-on.  
In the CubeX phase, the color-flavor and position space dependence of 
the condensate is given by
\begin{equation}
\begin{split}
\Delta_{CF}(x)_{\cf} =& \epsilon_{2\alpha\beta}\epsilon_{2ij} \, 
2\Delta \Biggl[ 
\cos \frac{2\pi}{a} \left( x+y+z\right) + \cos \frac{2\pi}{a}\left(-x-y+z\right) \Biggr]\\
&+ \epsilon_{3\alpha\beta}\epsilon_{3ij} \, 2\Delta \Biggl[
\cos \frac{2\pi}{a} \left( -x+y+z\right) + \cos \frac{2\pi}{a}\left(x-y+z\right) \Biggr]\ .
\label{CubeXStructure}
\end{split}
\end{equation}
We provide a depiction of this condensate in Fig.~\ref{CubeXContours}.

We are confident that 2Cube45z is the most favorable structure
obtained by rotating one cube relative to another. We are not as
confident that CubeX is the best possible structure with fewer than
8+8 vectors.  However, the two most favorable structures that we
have found, 2Cube45z and CubeX, are impressively robust and make the
case that three-flavor crystalline color superconducting phases are
the ground state of cold quark matter over a wide range of densities.
If even better crystal structures can be found, this will only further
strengthen this case.

\begin{figure}[t]
\bc
\includegraphics[width=0.6\hsize,angle=0]{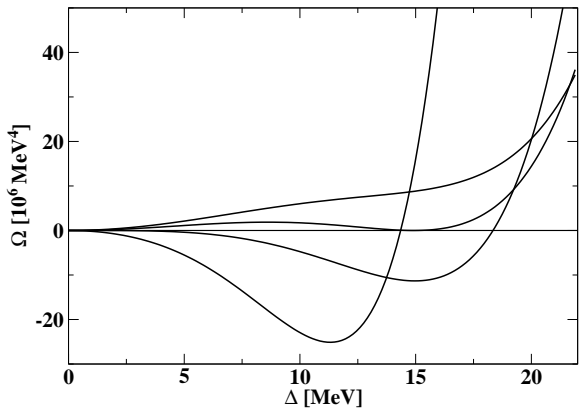}
\ec
\caption{
Free energy $\Omega$ vs.~$\Delta$ for the CubeX crystal structure,
described in Section \ref{sec:LOFFmultiple}, at four values of
$M_s^2/\mu$. From top curve to bottom curve, as judged from the left
half of the figure, the curves are $M_s^2/\mu=240$, $218.61$, $190$,
and $120$~MeV, corresponding to $\alpha=0.233$, $0.140$, 0, $-0.460$.
The first order phase transition occurs at $M_s^2/\mu=218.61$~MeV.
The values of $\Delta$ and $\Omega$ at the minima of curves like these
are what we plot in Figs.~\ref{deltavsx} and \ref{omegavsx}.
}
\label{OmegaVsDeltaFig}
\end{figure}

\subsection{Free energy comparisons and conclusions}
\label{sec:LOFFfree}

The gap parameter $\Delta$ and free energy
$\Omega(\Delta)$ can be evaluated for all the crystal structures whose 
Ginzburg-Landau coefficients have been determined.\cite{Rajagopal:2006ig}
For a given
crystal structure, $\Omega(\Delta)$ is given by Eq.~(\ref{effective free energy}), 
with $\bar\beta_{\rm eff}$ and $\bar\gamma_{\rm eff}$ taken 
from Table II of Ref.~\refcite{Rajagopal:2006ig}.
The quadratic coefficient $\alpha$ is related to $\delta\mu$ by Eq.~(\ref{AlphaEqn2}).
Recall that we have made the approximation that
$\delta\mu_2=\delta\mu_3=\delta\mu=M_s^2/(8\mu)$, valid up to corrections
of order $M_s^3/\mu^4$.   At any value of $M_s^2/\mu$, we can
evaluate $\alpha(\delta\mu)$ and hence $\Omega(\Delta)$, 
determine $\Delta$ by minimizing $\Omega$,
and finally evaluate the free energy $\Omega$ at the minimum.  
In Fig.~\ref{OmegaVsDeltaFig},
we give an example of $\Omega(\Delta)$ for various $M_s^2/\mu$
for one crystal structure with a first
order phase transition (CubeX), illustrating how the first order phase
transition is found, and how the $\Delta$ solving the gap
equations --- i.e. minimizing $\Omega$ --- is found.
In Figs.~\ref{deltavsx} and \ref{omegavsx},
we plot $\Delta$
and $\Omega$ at the minimum versus $M_s^2/\mu$ 
for the two most favorable crystal structures that we have found, namely the
CubeX and 2Cube45z described in Section \ref{sec:LOFFmultiple}.

In Figs.~\ref{deltavsx} and \ref{omegavsx}, we have chosen
the interaction strength between quarks such that the CFL gap parameter at $M_s=0$
is $\Delta_0=25$~MeV.  However, our results for both the
gap parameters and the condensation energy for any
of the crystalline phases can easily be scaled to any value of
$\Delta_0$.  Recall that the quartic and sextic coefficients in
the Ginzburg-Landau free energy do not depend on $\Delta_0$.  
And, recall from Eq.~(\ref{AlphaEqn2}) that
$\Delta_0$ enters $\alpha$ only through the combination $\Delta_{\rm 2SC}/\delta\mu$,
where $\Delta_{\rm 2SC}=2^{\frac{1}{3}}\Delta_0$ and $\delta\mu= M_s^2/(8\mu)$.
This means that if we pick a $\Delta_0\neq 25$~MeV, the 
curves describing the gap parameters for the
crystalline phases in Fig.~\ref{deltavsx} are precisely unchanged
if we rescale both the vertical and horizontal axes proportional to
$\Delta_0/25$~MeV.  In the case of Fig.~\ref{omegavsx}, the vertical
axis must be rescaled by $(\Delta_0/25~{\rm MeV})^2$.
Of course, the weak-coupling
approximation $\Delta_0\ll\mu$, which we have used to simplify
the calculation of the Ginzburg-Landau coefficients 
in Ref.~\refcite{Rajagopal:2006ig}
will break down if we scale $\Delta_0$ to be too large.
We cannot evaluate up to what $\Delta_0$ we can scale our results
reliably without doing a calculation that goes beyond the weak-coupling
limit. However, such calculations have been done for the gCFL phase 
in Ref.~\refcite{Fukushima:2004zq},
where it turns out that the gaps and
condensation energies plotted in Figs.~\ref{deltavsx} and \ref{omegavsx} 
scale with $\Delta_0$ and $\Delta_0^2$ to good accuracy 
for $\Delta_0\leq 40$~MeV with $\mu=500$~MeV, but the scaling is
significantly less accurate for $\Delta_0=100$~MeV.
Of course, for $\Delta_0$ as large as $100$~MeV, any quark matter in
a compact star is likely to be in the CFL phase.  Less symmetrically
paired quark matter, which our results suggest is in a crystalline
color superconducting phase, will occur in compact stars only
if $\Delta_0$ is smaller, in the range where our results can be 
expected to scale well.

Fig.~\ref{deltavsx} can be used to evaluate the validity of the
Ginzburg-Landau approximation.  The simplest criterion is to compare
the $\Delta$'s for the crystalline phases to the CFL gap parameter
$\Delta_0$.  This is the correct criterion in the vicinity of the 2nd
order phase transition point, where $\delta\mu = M_s^2/(8\mu) \approx
\Delta_0$ as in (\ref{AlphaZero}).  Well to the left, it is more
appropriate to compare the $\Delta$'s for the crystalline phase to
$\delta\mu=M_s^2/(8\mu)$.  By either criterion, the Ginzburg-Landau
approximation is at the edge of its domain of validity, a result which
we expected based on the general arguments of Section
\ref{sec:LOFFgeneral}.  Therefore, although we expect that the
qualitative lessons that we have learned about the favorability of
crystalline phases in three-flavor quark matter are valid, and expect
that the relative favorability of the 2Cube45z and CubeX structures
and the qualitative size of their $\Delta$ and condensation energy are
trustworthy, we do not expect quantitative reliability of our results.
There is therefore strong motivation to analyze crystalline color
superconducting quark matter with these two crystal structures without
making a Ginzburg-Landau approximation.  It will be very interesting
to see whether the Ginzburg-Landau approximation underestimates
$\Delta$ and the condensation energy for the crystalline phases with
CubeX and 2Cube45z crystal structures, as it does for the much simpler
structure in which $\Delta_2(\rr)$ and $\Delta_3(\rr)$ are each single
plane waves.\cite{Mannarelli:2006fy}

Fig.~\ref{omegavsx} makes it clear that  
the three-flavor crystalline color superconducting
phases with the two most favored crystal structures that have been
found are robust by any measure.  Their condensation energies 
reach about half that of  the CFL phase at $M_s=0$. 
Correspondingly, these two crystal structures are favored over the wide
range of $M_s^2/\mu$ seen in Fig.~\ref{omegavsx} and
given in Eq.~(\ref{WideLOFFWindow}).

Taken literally, Fig.~\ref{omegavsx} indicates that within the regime
(\ref{WideLOFFWindow}) of the phase diagram occupied by crystalline
color superconducting quark matter, the 2Cube45z phase is favored at
lower densities and the CubeX phase is favored at higher densities.
Although we do have
qualitative arguments why 2Cube45z and CubeX are favored over other phases,
we have no qualitative argument why one should be favored over the other.
Also, in this context the Ginzburg-Landau approximation is not reliable
enough for us to conclude that one phase is favored at higher
densities while the other is favored at lower ones. 
The proper conclusion from these results is that 2Cube45z and CubeX
are the two most favorable phases we have
found, that both are robust, that the crystalline color superconducting phase
of three-flavor quark matter with one crystal structure or the other occupies
a wide swath of the QCD phase diagram,
and that their free energies are similar enough to each other that
a definitive comparison of their free energies will require a calculation that
goes beyond the Ginzburg-Landau approximation that we have used here.

\subsection{Implications and Future Work}

The purpose of this calculation has been to find candidates for the
true ground state of quark matter in the ``gCFL window'', the 
range of values of $M_s^2/\mu$ where calculations that are limited to 
isotropic phases predict that the gCFL phase is favored.
Fig.~\ref{omegavsx} shows that, according to a 
Ginzburg-Landau theory keeping terms up to sextic order, in most of this
range the gCFL phase can be
replaced by a much more favorable three-flavor crystalline color
superconducting phase and that these crystalline phases may
be favored over a wide range of $M_s^2/\mu$ above the gCFL window also.
Fig.~\ref{omegavsx} also indicates that at the lower end of the
range of $M_s^2/\mu$, closest to the
CFL$\rightarrow$gCFL transition, there is a narrow region where it is hard
to find a crystalline color superconducting phase
with lower free energy than the gCFL phase.
This narrow region is thus the most likely place
to find a ground state consisting of the gCFL phase augmented
by current-carrying meson condensates described in 
Refs.~\refcite{Kryjevski:2005qq,Gerhold:2006dt}.
Except within this window, the crystalline color superconducting
phases with either the CubeX or the 2Cube45z crystal structure
provide an attractive resolution to the instability of the gCFL phase.

Our Ginzburg-Landau calculation provides reason to speculate that most
or possibly {\em all} of the quark matter core of a compact star might
be in a crystalline color superconducting phase. For example,
substituting the quite reasonable values $\Delta_0=25$~MeV and
$M_s=250$~MeV into Eq.~(\ref{WideLOFFWindow}) we find that the
crystalline phases are favored over the range $240 {\rm MeV} < \mu <
847 {\rm MeV}$.  Obviously at the lower density end of this range all
quark matter phases are replaced by nuclear matter, and the high end
of this range extends far beyond the values $\mu\sim 500$~MeV that are
expected in the center of compact stars. We therefore find that with
these choices for the parameters the crystalline phase covers the
entire range of expected densities of stellar quark matter.  Of
course, if $\Delta_0$ is larger, say $\sim 100$~MeV, the entire quark
matter core could be in the CFL phase. And, there are reasonable
values of $\Delta_0$ and $M_s$ for which the outer layer of a possible
quark matter core would be in a crystalline phase while the inner core
would not.  We do not know $\Delta_0$ and $M_s$ well enough to predict
what phases of quark matter occur in compact stars. However, from our
results it is clear that one of the possible configurations of a
compact star, ultimately to be confirmed or refuted by astrophysical
observations, would involve a quark matter core that is mostly or
entirely in a crystalline color superconducting phase.

Now that we have two well-motivated ans\"atze
for the crystal structure of three-flavor crystalline
color superconducting quark matter and a good qualitative
guide to the scale of $\Delta$ and $\Omega(\Delta)$, 
we should be able to make progress
toward the calculation of astrophysically observable preperties
of compact stars.
For example, we would like to predict the effects of a 
crystalline color superconducting quark core on the rate at
which neutron stars cool by neutrino emission and 
on the occurrence and phenomenology of pulsar glitches.

For cooling, the relevant microphysical quantities are 
the specific heat and the
neutrino emissivity. The specific heat of the crystalline phases
will not be dramatically different from that of
unpaired quark matter: it rises linearly with 
$T$ because of the presence of
gapless quark excitations at the boundaries of the regions in momentum
space where there are unpaired quarks.\cite{Casalbuoni:2003sa}
(In contrast, in the gCFL case the heat capacity is parametrically
enhanced.\cite{Alford:2004zr})
However, the neutrino emissivity should turn out to be significantly
suppressed relative to that in unpaired quark matter. 
The evaluation of the phase
space for direct Urca neutrino emission from the CubeX and 2Cube45z phases
will be a nontrivial calculation, given that 
thermally excited gapless quarks 
occur only on patches of the Fermi surfaces, separated by the (many)
pairing rings. (The direct Urca processes $u + e \rightarrow s + \nu$
and $s\rightarrow  u + e + \bar\nu$ 
require $s$, $u$ and $e$ to all be within $T$
of a place in momentum space where they are  gapless and at the same
time to have ${\bf p}_u + {\bf p}e = {\bf p}_s$ to within $T$.  
Here, $T\sim$~keV is
very small compared to all the scales relevant to the description of
the crystalline phase itself.)

The idea that a region of a star made of quark matter in a
crystalline phase could be the place where (some) glitches 
originate goes back to
Ref.~\refcite{Alford:2000ze}, where it was suggested that a crystalline 
color superconductor condensate
could pin rotational vortices. This leads to the
possibility that the presence of crystalline
quark matter in compact stars could be strongly constrained by
comparing their predicted glitch phenomenology with what is actually observed.

There are two key microphysical properties of crystalline quark matter
that must be estimated before glitch phenomenology can be addressed:
the pinning force on rotational vortices and the shear modulus
of the crystal.  In order for glitches to occur it is necessary for
rotational vortices in some region of the star to be pinned and
immobile while the spinning pulsar's angular velocity slows over
years.  But this can only occur if the pinning force is strong enough
to lock the vortices to the crystal structure {\em and} if the
structure resists deformation, i.e.~has a large enough shear modulus.
(The glitches themselves are then triggered by the catastrophic
unpinning and motion of long-immobile vortices.)

To estimate the pinning force we must analyze
how the CubeX and 2Cube45z phases respond when rotated. We expect
vortices to form and be pinned at the intersections
of the nodal planes at which condensates vanish. This analysis
faces two complications. Firstly,
baryon number current can be carried by gradients
in the phase of either the $\langle us \rangle$
crystalline condensate or the $\langle ud \rangle$ condensate or both,
so we must determine which types of vortices are formed.
Secondly, the vortex core size, $1/\Delta$, is only a factor
of three to four smaller than the lattice spacing $a$.  This means that
the vortices cannot be thought of as pinned by an unchanged crystal;
the vortices themselves will qualitatively deform the crystalline condensate in
their vicinity.  

The second microphysical quantity that is required is the shear
modulus of the crystal.  This can be related to the coefficients in
the low energy effective theory that describes the phonon modes of the
crystal.\cite{Casalbuoni:2001gt,Casalbuoni:2002pa,Casalbuoni:2002my}
This effective theory has been analyzed, and its coefficients
calculated, for the two-flavor crystalline color superconductor with
face-centered cubic symmetry.\cite{Casalbuoni:2002my}  Analyzing
the phonons in the three-flavor crystalline color superconducting phases with
the 2Cube45z and CubeX crystal structures is thus a priority for future
work.

The instability of the gCFL phase was a significant bend in
the road to understanding the properties of dense matter, but one that is now
receding behind us. With the emergence of two robust crystalline phases,
possibly accounting for as much as all of the quark matter that may occur, 
we can glimpse on the
road ahead the microphysical calculations prerequisite to 
identifying observations that 
can be used to rule out (or in) the presence of 
quark matter
in a crystalline color superconducting phase within neutron stars.

\section{Coda}

The project of delineating a plausible phase diagram for high-density
quark matter is still not complete. We have discussed some ideas for
the ``non-CFL'' region, but there are others such as a suggested gluon
condensation in two-flavor quark matter,\cite{Gorbar:2005rx} and
deformation of the Fermi surfaces (discussed so far only in
non-beta-equilibrated nuclear matter\cite{Sedrakian:2003tr}).  It is
very interesting to note that the problem of how a system with pairing
responds to a stress that separates the chemical potentials of the
pairing species is a very generic one, arising in condensed matter
systems and cold atom systems as well as in quark matter.  Recent work
by Son and Stephanov\cite{Son:2005qx} and Mannarelli, Nardulli and
Ruggieri\cite{Mannarelli:2006hr} on a two-species model characterized
by a diluteness parameter and a splitting potential shows that between
the BCS-paired region and the unpaired region in the phase diagram one
should expect a translationally-broken region. In QCD this could
correspond to a $p$-wave meson condensate or a crystalline color
superconducting state. What is particularly exciting is that the
technology of cold atom traps has advanced to the point where fermion
superfluidity can now be seen in conditions where many of the
important parameters can be manipulated, and it may soon be possible
to investigate the response of the pairing to external stress under
controlled experimental conditions.  Pairing in a gas of ultracold
fermionic atoms can never yield a complete analogue of pairing in
dense quark matter, because the atoms are neutral and thus the
requirement of neutrality imposes no constraints, whereas we have seen
that in the quark matter context imposing both color and electric
neutrality is a crucial qualitative driver for the physics, for
example precluding phase separation which can easily happen in the
cold atom system.  However, even without yielding a complete analogue
these experiments could answer qualitative questions of interest. For
example, should they discover a translationally broken regime in the
absence of rotation, it will be very interesting first to see what
crystal structure emerges and second to study the response of such a
phase to rotation.



\vspace{0.2in}

KR acknowledges the
hospitality of the Nuclear Theory Group at LBNL. This research was
supported in part by the Office of Nuclear Physics of the Office of
Science of the U.S.~Department of Energy under contracts
\#DE-AC02-05CH11231, \#DE-FG02-91ER50628,
\#DE-FG01-04ER0225 (OJI) and cooperative research agreement
\#DF-FC02-94ER40818.




\begin{thebibliography}{99}

\bibitem{BCS}
  J.~Bardeen, L.~Cooper, J.~Schrieffer, Phys. Rev. {\bf 106}, 162 (1957);
  Phys. Rev. {\bf 108}, 1175 (1957)

\bibitem{Alford:1998mk}
M.~Alford, K.~Rajagopal and F.~Wilczek,
Nucl.\ Phys.\  {\bf B537}, 443 (1999)
\mbox{[hep-ph/9804403]}.

\bibitem{Reviews}
For reviews, see
K.~Rajagopal and F.~Wilczek,
arXiv:hep-ph/0011333;
M.~G.~Alford,
Ann.\ Rev.\ Nucl.\ Part.\ Sci.\  {\bf 51}, 131 (2001)
[arXiv:hep-ph/0102047];
  D.~K.~Hong,
  Acta Phys.\ Polon.\ B {\bf 32}, 1253 (2001)
  [hep-ph/0101025].
G.~Nardulli,
Riv.\ Nuovo Cim.\  {\bf 25N3}, 1 (2002)
[arXiv:hep-ph/0202037];
S.~Reddy,
Acta Phys.\ Polon.\ B {\bf 33}, 4101 (2002)
[arXiv:nucl-th/0211045];
T.~Sch\"afer,
arXiv:hep-ph/0304281;
D.~H.~Rischke, Prog.\ Part.\ Nucl.\ Phys.\ {\bf 52}, 197 (2004)
[arXiv:nucl-th/0305030];
  M.~Alford,
  Prog.\ Theor.\ Phys.\ Suppl.\  {\bf 153}, 1 (2004)
  [arXiv:nucl-th/0312007];
  M.~Buballa,
  Phys.\ Rept.\  {\bf 407}, 205 (2005)
  [arXiv:hep-ph/0402234];
  H.~c.~Ren,
  arXiv:hep-ph/0404074;
I. Shovkovy, arXiv:nucl-th/0410091;
T. Sch\"afer, arXiv:hep-ph/0509068;
T. Sch\"afer, arXiv:hep-ph/0602067.



\bibitem{Alford:2003fq}
M.~Alford, C.~Kouvaris and K.~Rajagopal,
Phys.\ Rev.\ Lett.\  {\bf 92}, 222001 (2004)
[arXiv:hep-ph/0311286].

\bibitem{Alford:2004hz}
  M.~Alford, C.~Kouvaris and K.~Rajagopal,
  Phys.\ Rev.\ D {\bf 71}, 054009 (2005)
  [arXiv:hep-ph/0406137].

\bibitem{Alford:2004zr}
M.~Alford, P.~Jotwani, C.~Kouvaris, J.~Kundu and K.~Rajagopal,
arXiv:astro-ph/0411560.

\bibitem{Huang:2004bg}
M.~Huang and I.~A.~Shovkovy,
Phys. Rev. D {\bf 70}, 051501 (2004)
[arXiv:hep-ph/0407049];
  M.~Huang and I.~A.~Shovkovy,
  Phys.\ Rev.\ D {\bf 70}, 094030 (2004)
  [arXiv:hep-ph/0408268].


\bibitem{Casalbuoni:2004tb}
  R.~Casalbuoni, R.~Gatto, M.~Mannarelli, G.~Nardulli and M.~Ruggieri,
  Phys.\ Lett.\ B {\bf 605}, 362 (2005)
  [Erratum-ibid.\ B {\bf 615}, 297 (2005)]
  [arXiv:hep-ph/0410401].

\bibitem{Giannakis:2004pf}
  I.~Giannakis and H.~C.~Ren,
  Phys.\ Lett.\ B {\bf 611}, 137 (2005)
  [arXiv:hep-ph/0412015].


\bibitem{Alford:2005qw}
  M.~Alford and Q.~h.~Wang,
  J.\ Phys.\ G {\bf 31}, 719 (2005)
  [arXiv:hep-ph/0501078].

\bibitem{Huang:2005pv}
  M.~Huang,
  Phys.\ Rev.\ D {\bf 73}, 045007 (2006)
  [arXiv:hep-ph/0504235].

\bibitem{Fukushima:2005cm}
  K.~Fukushima,
  Phys.\ Rev.\ D {\bf 72}, 074002 (2005)
  [arXiv:hep-ph/0506080].

\bibitem{Gorbar:2005rx}
  E.~V.~Gorbar, M.~Hashimoto and V.~A.~Miransky,
  Phys.\ Lett.\ B {\bf 632}, 305 (2006)
  [arXiv:hep-ph/0507303].

\bibitem{Kryjevski:2005qq}
  A.~Kryjevski,
  arXiv:hep-ph/0508180;
  T.~Schafer,
  Phys.\ Rev.\ Lett.\  {\bf 96}, 012305 (2006)
  [arXiv:hep-ph/0508190].

\bibitem{Gorbar:2006up}
  E.~V.~Gorbar, M.~Hashimoto, V.~A.~Miransky and I.~A.~Shovkovy,
  arXiv:hep-ph/0602251.

\bibitem{Iida:2006df}
  K.~Iida and K.~Fukushima,
  arXiv:hep-ph/0603179.

\bibitem{Fukushima:2006su}
  K.~Fukushima,
  Phys.\ Rev.\ D {\bf 73}, 094016 (2006)
  [arXiv:hep-ph/0603216].

\bibitem{Gerhold:2006dt}
  A.~Gerhold and T.~Schafer,
  arXiv:hep-ph/0603257.



\bibitem{Alford:2000ze}
  M.~G.~Alford, J.~A.~Bowers and K.~Rajagopal,
  Phys.\ Rev.\ D {\bf 63}, 074016 (2001)
  [arXiv:hep-ph/0008208].

  
\bibitem{Alford:1999pa}
  M.~G.~Alford, J.~Berges and K.~Rajagopal,
  Nucl.\ Phys.\ B {\bf 558}, 219 (1999)
  [arXiv:hep-ph/9903502];
  T.~Schafer and F.~Wilczek,
  Phys.\ Rev.\ D {\bf 60}, 074014 (1999)
  [arXiv:hep-ph/9903503].

\bibitem{Schafer:1999fe}
  T.~Schafer,
  Nucl.\ Phys.\ B {\bf 575}, 269 (2000)
  [arXiv:hep-ph/9909574].


\bibitem{Shovkovy:1999mr}
  I.~A.~Shovkovy and L.~C.~R.~Wijewardhana,
  Phys.\ Lett.\ B {\bf 470}, 189 (1999)
  [arXiv:hep-ph/9910225].

\bibitem{Evans:1999at}
  N.~J.~Evans, J.~Hormuzdiar, S.~D.~H.~Hsu and M.~Schwetz,
  Nucl.\ Phys.\ B {\bf 581}, 391 (2000)
  [arXiv:hep-ph/9910313].


\bibitem{Pisarski:1999cn}
  R.~D.~Pisarski and D.~H.~Rischke,
nucl-th/9907094.


\bibitem{IwaIwa}
  M.~Iwasaki, T.~Iwado, Phys. Lett. {\bf B350}, 163 (1995);
  M.~Iwasaki, Prog. Theor. Phys. Suppl. {\bf 120}, 187 (1995)

\bibitem{Schafer:2000tw}
  T.~Schafer,
  Phys.\ Rev.\ D {\bf 62}, 094007 (2000)
  [arXiv:hep-ph/0006034].

\bibitem{Buballa:2002wy}
  M.~Buballa, J.~Hosek and M.~Oertel,
  Phys.\ Rev.\ Lett.\  {\bf 90}, 182002 (2003)
  [arXiv:hep-ph/0204275].

\bibitem{Alford:2002rz}
  M.~G.~Alford, J.~A.~Bowers, J.~M.~Cheyne and G.~A.~Cowan,
  Phys.\ Rev.\ D {\bf 67}, 054018 (2003)
  [arXiv:hep-ph/0210106].

\bibitem{Schmitt:2002sc}
  A.~Schmitt, Q.~Wang and D.~H.~Rischke,
  Phys.\ Rev.\ D {\bf 66}, 114010 (2002)
  [arXiv:nucl-th/0209050].

\bibitem{Schmitt:2004et}
  A.~Schmitt,
  Phys.\ Rev.\ D {\bf 71}, 054016 (2005)
  [arXiv:nucl-th/0412033];

\bibitem{Iida:2000ha}
  K.~Iida and G.~Baym,
  Phys.\ Rev.\ D {\bf 63}, 074018 (2001)
  [Erratum-ibid.\ D {\bf 66}, 059903 (2002)]
  [arXiv:hep-ph/0011229].


\bibitem{Amore:2001uf}
P.~Amore, M.~C.~Birse, J.~A.~McGovern and N.~R.~Walet,
Phys.\ Rev.\ D {\bf 65}, 074005 (2002)
[arXiv:hep-ph/0110267].

\bibitem{Alford:2002kj}
  M.~Alford and K.~Rajagopal,
  JHEP {\bf 0206}, 031 (2002)
  [arXiv:hep-ph/0204001].

\bibitem{Steiner:2002gx}
A.~W.~Steiner, S.~Reddy and M.~Prakash,
Phys.\ Rev.\ D {\bf 66}, 094007 (2002)
[arXiv:hep-ph/0205201].

\bibitem{Huang:2002zd}
  M.~Huang, P.~f.~Zhuang and W.~q.~Chao,
  Phys.\ Rev.\ D {\bf 67}, 065015 (2003)
  [arXiv:hep-ph/0207008];

\bibitem{Neumann:2002jm}
F.~Neumann, M.~Buballa and M.~Oertel,
Nucl.\ Phys.\ A {\bf 714}, 481 (2003)
[arXiv:hep-ph/0210078].

\bibitem{Bedaque:2001je}
P.~F.~Bedaque and T.~Sch\"afer,
Nucl.\ Phys.\ A {\bf 697}, 802 (2002)
[arXiv:hep-ph/0105150];
D.~B.~Kaplan and S.~Reddy,
Phys.\ Rev.\ D {\bf 65}, 054042 (2002)
[arXiv:hep-ph/0107265];
  A.~Kryjevski, D.~B.~Kaplan and T.~Sch\"afer,
  Phys.\ Rev.\ D {\bf 71}, 034004 (2005)
  [arXiv:hep-ph/0404290];

\bibitem{Kryjevski:2004jw}
  A.~Kryjevski and T.~Sch\"afer,
  Phys.\ Lett.\ B {\bf 606}, 52 (2005)
  [arXiv:hep-ph/0407329];
  A.~Kryjevski and D.~Yamada,
  Phys.\ Rev.\ D {\bf 71}, 014011 (2005)
  [arXiv:hep-ph/0407350].

\bibitem{Schafer:2002ty}
  T.~Schafer,
  Phys.\ Rev.\ D {\bf 65}, 094033 (2002)
  [arXiv:hep-ph/0201189].

\bibitem{Buballa:2004sx}
  M.~Buballa,
  Phys.\ Lett.\ B {\bf 609}, 57 (2005)
  [arXiv:hep-ph/0410397].

\bibitem{Forbes:2004ww}
  M.~M.~Forbes,
  Phys.\ Rev.\ D {\bf 72}, 094032 (2005)
  [arXiv:hep-ph/0411001].





\bibitem{Alford:2004nf}
  M.~Alford, C.~Kouvaris and K.~Rajagopal,
  arXiv:hep-ph/0407257.

\bibitem{Ruster:2004eg}
S.~B.~R\"uster, I.~A.~Shovkovy and D.~H.~Rischke,
Nucl.\ Phys.\ A {\bf 743}, 127 (2004)
[arXiv:hep-ph/0405170].

\bibitem{Fukushima:2004zq}
  K.~Fukushima, C.~Kouvaris and K.~Rajagopal,
  Phys.\ Rev.\ D {\bf 71}, 034002 (2005)
  [arXiv:hep-ph/0408322].


\bibitem{Abuki:2004zk}
  H.~Abuki, M.~Kitazawa and T.~Kunihiro,
  Phys.\ Lett.\ B {\bf 615}, 102 (2005)
  [arXiv:hep-ph/0412382].

\bibitem{Ruster:2005jc}
  S.~B.~R\"uster, V.~Werth, M.~Buballa, I.~A.~Shovkovy and D.~H.~Rischke,
  Phys.\ Rev.\ D {\bf 72}, 034004 (2005)
  [arXiv:hep-ph/0503184].

\bibitem{Blaschke:2005uj}
  D.~Blaschke, S.~Fredriksson, H.~Grigorian, A.~M.~Oztas and F.~Sandin,
  Phys.\ Rev.\ D {\bf 72}, 065020 (2005)
  [arXiv:hep-ph/0503194].

\bibitem{Shovkovy:2003uu}
I.~Shovkovy and M.~Huang,
Phys.\ Lett.\ B {\bf 564}, 205 (2003)
[arXiv:hep-ph/0302142];

\bibitem{Huang:2003xd}
  M.~Huang and I.~Shovkovy,
  Nucl.\ Phys.\ A {\bf 729}, 835 (2003)
  [arXiv:hep-ph/0307273].

\bibitem{Gubankova:2003uj}
E.~Gubankova, W.~V.~Liu and F.~Wilczek,
Phys.\ Rev.\ Lett.\  {\bf 91}, 032001 (2003)
[arXiv:hep-ph/0304016].

\bibitem{Rajagopal:2000ff}
K.~Rajagopal and F.~Wilczek,
Phys.\ Rev.\ Lett.\  {\bf 86}, 3492 (2001)
[arXiv:hep-ph/0012039].

\bibitem{Rajagopal:2006ig}
  K.~Rajagopal and R.~Sharma,
  arXiv:hep-ph/0605316.

\bibitem{Bedaque:2003hi}
  P.~F.~Bedaque, H.~Caldas and G.~Rupak,
  Phys.\ Rev.\ Lett.\  {\bf 91}, 247002 (2003)
  [arXiv:cond-mat/0306694];
  M.~M.~Forbes,  E.~Gubankova, W.~V.~Liu and F.~Wilczek,
  Phys.\ Rev.\ Lett.\  {\bf 94}, 017001 (2005)
  [arXiv:hep-ph/0405059];
  J.~Carlson and S.~Reddy,
  Phys.\ Rev.\ Lett.\  {\bf 95}, 060401 (2005)
  [arXiv:cond-mat/0503256].

\bibitem{KetterleImbalancedSpin} M.W.~Zwierlein,
A.~Schirotzek, C.H.~Schunck, and W.~Ketterle, Science {\bf 311},
492 (2006) [arXiv:cond-mat/0511197].

\bibitem{HuletPhaseSeparation} G.B.~Partridge, W.~Li, R.I.~Kamar,
Y.-a.~Liao, and R.G.~Hulet,
 Science {\bf 311}, 503 (2006) [arXiv:cond-mat/0511752].

\bibitem{Reddy:2004my}
 S.~Reddy and G.~Rupak,
  Phys.\ Rev.\ C {\bf 71}, 025201 (2005)
  [arXiv:nucl-th/0405054].

\bibitem{Bailin:1983bm}
  D.~Bailin and A.~Love,
  Phys.\ Rept.\  {\bf 107}, 325 (1984).

\bibitem{Alford:1997zt}
  M.~G.~Alford, K.~Rajagopal and F.~Wilczek,
  Phys.\ Lett.\ B {\bf 422}, 247 (1998)
  [arXiv:hep-ph/9711395].

\bibitem{Rapp:1997zu}
  R.~Rapp, T.~Sch\"afer, E.~V.~Shuryak and M.~Velkovsky,
  Phys.\ Rev.\ Lett.\  {\bf 81}, 53 (1998)
  [arXiv:hep-ph/9711396].

\bibitem{Rajagopal:2005dg}
  K.~Rajagopal and A.~Schmitt,
  Phys.\ Rev.\ D {\bf 73}, 045003 (2006)
  [arXiv:hep-ph/0512043].

\bibitem{Bowers:2001ip}
  J.~A.~Bowers, J.~Kundu, K.~Rajagopal and E.~Shuster,
  Phys.\ Rev.\ D {\bf 64}, 014024 (2001)
  [arXiv:hep-ph/0101067].
  
\bibitem{Casalbuoni:2001gt}
  R.~Casalbuoni, R.~Gatto, M.~Mannarelli and G.~Nardulli,
  Phys.\ Lett.\ B {\bf 511}, 218 (2001)
  [arXiv:hep-ph/0101326].

\bibitem{Leibovich:2001xr}
A.~K.~Leibovich, K.~Rajagopal and E.~Shuster, Phys.\ Rev.\ D {\bf 64},
094005 (2001) [arXiv:hep-ph/0104073].

\bibitem{Kundu:2001tt}
J.~Kundu and K.~Rajagopal,
Phys.\ Rev.\ D {\bf 65}, 094022 (2002)
[arXiv:hep-ph/0112206].

\bibitem{Bowers:2002xr}
J.~A.~Bowers and K.~Rajagopal,
Phys.\ Rev.\ D {\bf 66}, 065002 (2002)
[arXiv:hep-ph/0204079].

\bibitem{Casalbuoni:2003wh}
R.~Casalbuoni and G.~Nardulli,
Rev.\ Mod.\ Phys.\  {\bf 263}, 320 (2004)
[arXiv:hep-ph/0305069].

\bibitem{Casalbuoni:2003sa}
  R.~Casalbuoni, R.~Gatto, M.~Mannarelli, G.~Nardulli, M.~Ruggieri and S.~Stramaglia,
  Phys.\ Lett.\ B {\bf 575}, 181 (2003)
  [Erratum-ibid.\ B {\bf 582}, 279 (2004)]
  [arXiv:hep-ph/0307335].


\bibitem{Casalbuoni:2004wm}
  R.~Casalbuoni, M.~Ciminale, M.~Mannarelli, G.~Nardulli, M.~Ruggieri and
R.~Gatto,
  Phys.\ Rev.\ D {\bf 70}, 054004 (2004)
  [arXiv:hep-ph/0404090].

\bibitem{Casalbuoni:2005zp}
  R.~Casalbuoni, R.~Gatto, N.~Ippolito, G.~Nardulli and M.~Ruggieri,
  Phys.\ Lett.\ B {\bf 627}, 89 (2005)
  [arXiv:hep-ph/0507247].

\bibitem{Ciminale:2006sm}
  M.~Ciminale, G.~Nardulli, M.~Ruggieri and R.~Gatto,
  Phys.\ Lett.\ B {\bf 636}, 317 (2006)
  [arXiv:hep-ph/0602180].

\bibitem{Mannarelli:2006fy}
  M.~Mannarelli, K.~Rajagopal and R.~Sharma,
  arXiv:hep-ph/0603076.

\bibitem{LOFF}
A.~I.~Larkin and Yu.~N.~Ovchinnikov, Zh. Eksp. Teor. Fiz.~{\bf 47}, 1136
(1964)[Sov. Phys. JETP {\bf 20}, 762 (1965)];
P.~Fulde and R.~A.~Ferrell, Phys.\ Rev.\ {\bf 135}, A550 (1964);
S.~Takada and T.~Izuyama, Prog.\  Theor.\ Phys.\ {\bf 41}, 635 (1969).

\bibitem{Hong:2005jy}
  D.~K.~Hong,
  arXiv:hep-ph/0506097.

\bibitem{Alford:2005kj}
  M.~Alford and Q.~h.~Wang,
  J.\ Phys.\ G {\bf 32}, 63 (2006)
  [arXiv:hep-ph/0507269].

\bibitem{Alford:2005yy}
  M.~G.~Alford and G.~A.~Cowan,
  J.\ Phys.\ G {\bf 32}, 511 (2006)
  [arXiv:hep-ph/0512104].

\bibitem{Alford:2001zr}
  M.~G.~Alford, K.~Rajagopal, S.~Reddy and F.~Wilczek,
  Phys.\ Rev.\ D {\bf 64}, 074017 (2001)
  [arXiv:hep-ph/0105009].

\bibitem{Gerhold:2003js}
A.~Gerhold and A.~Rebhan,
Phys.\ Rev.\ D {\bf 68}, 011502 (2003)
[arXiv:hep-ph/0305108];
A.~Kryjevski,
Phys.\ Rev.\ D {\bf 68}, 074008 (2003)
[arXiv:hep-ph/0305173];
  A.~Gerhold,
  Phys.\ Rev.\ D {\bf 71}, 014039 (2005)
  [arXiv:hep-ph/0411086];
D.~D.~Dietrich and D.~H.~Rischke,
Prog.\ Part.\ Nucl.\ Phys.\  {\bf 53}, 305 (2004)
[arXiv:nucl-th/0312044];

\bibitem{Casalbuoni:2002pa}
  R.~Casalbuoni, R.~Gatto, M.~Mannarelli and G.~Nardulli,
  Phys.\ Rev.\ D {\bf 66}, 014006 (2002)
  [arXiv:hep-ph/0201059].

\bibitem{Casalbuoni:2002my}
  R.~Casalbuoni, E.~Fabiano, R.~Gatto, M.~Mannarelli and G.~Nardulli,
  Phys.\ Rev.\ D {\bf 66}, 094006 (2002)
  [arXiv:hep-ph/0208121].

 \bibitem{Sedrakian:2003tr}
  A.~Sedrakian,
  arXiv:nucl-th/0312053.

\bibitem{Son:2005qx}
  D.~T.~Son and M.~A.~Stephanov,
  arXiv:cond-mat/0507586.

 \bibitem{Mannarelli:2006hr}
  M.~Mannarelli, G.~Nardulli and M.~Ruggieri,
  arXiv:cond-mat/0604579.
 
\end{thebibliography}
\end{document}